\documentclass[journal,12pt,draftclsnofoot,onecolumn]{IEEEtran}
\usepackage{mathpple}
\usepackage{times}
\usepackage{dsfont}
\usepackage{subcaption}
\usepackage{amsmath}  
\usepackage{amssymb}  
\usepackage{mathrsfs} 
\usepackage{theorem}  
\usepackage{cite}     
\usepackage{comment}  
\usepackage[colorlinks,
			linkcolor=black,
			anchorcolor=black,
			citecolor=black
			]{hyperref}
\usepackage{upref}
\usepackage{amsfonts}
\usepackage{dsfont}
\usepackage{verbatim}
\usepackage{algorithm}
\usepackage{algorithmic}
\usepackage[dvipsnames,usenames]{color}
\usepackage{enumerate}

\usepackage{graphicx}

\usepackage{latexsym}

\usepackage{color}
\usepackage{footnote}
\usepackage{booktabs}
\usepackage{cite}
\usepackage{hhline}
\usepackage{gensymb}
\usepackage{graphicx}
\usepackage{xcolor}
\usepackage{listings}
\usepackage[
top    = 0.75in,
bottom = 1.05in,
left   = 1in,
right  =1in
]{geometry}
\usepackage{caption}
\usepackage{romannum}
\usepackage{setspace}
\usepackage{bm}
\usepackage{tikz}
\usetikzlibrary{dsp}
\usetikzlibrary{graphs}

\usepackage{bbm}

\usepackage{multirow}

\usepackage{makecell}

\usepackage{pifont}
\newcommand{\RNum}[1]{\textup{\uppercase\expandafter{\romannumeral#1\relax}}}

\newcommand{\be}[1]{\begin{equation}\label{#1}}
\newcommand{\ee}{\end{equation}}

\newcommand{\floor}[1]{\lfloor{#1}\rfloor}

\newcommand{\qed}{\hfill$\Box$\\[1ex]}

\newcommand{\Cref}[1]{Co\-rol\-la\-ry\,\ref{#1}}

\theoremstyle{plain} \theorembodyfont{\normalfont\slshape}

\newtheorem{thm}{Theorem$\!$}

\newtheorem{prop}[thm]{Proposition$\!$}

\newtheorem{lem}[thm]{Lemma$\!$}

\newtheorem{cor}[thm]{Corollary$\!$}

\newtheorem{prob}[thm]{Problem$\!$}

\newtheorem{defi}[thm]{Definition$\!$}

\theorembodyfont{\normalfont}

\newtheorem{exam}{Example$\!$}
\newenvironment{example}{\begin{exam}\hspace*{-1ex}{\bf .}}{\end{exam}}

\newtheorem{remrk}{Remark$\!$}
\newenvironment{remark}{\begin{remrk}\hspace*{-1ex}{\bf .}}{\end{remrk}}

\definecolor{Codecolor}{named}{White}  

\begin{document}
\title{PR-NN: RNN-based Detection for \\ Coded Partial-Response Channels}

\author{\large Simeng Zheng,~\IEEEmembership{Student Member,~IEEE,} Yi Liu,~\IEEEmembership{Student Member,~IEEE,} \\ \IEEEauthorblockN{and Paul H. Siegel,~\IEEEmembership{Life Fellow,~IEEE}} \\
\thanks{
The authors are with the Center for Memory and Recording Research, Department of Electrical and Computer Engineering, University of California San Diego, La Jolla, CA 92093, USA (e-mail: sizheng@ucsd.edu; yil333@ucsd.edu; psiegel@ucsd.edu).
}
}
 
\maketitle
\pagenumbering{arabic}

\begin{abstract}

In this paper, we investigate the use of recurrent neural network (RNN)-based detection of magnetic recording channels with inter-symbol interference (ISI). We refer to the proposed detection method, which is intended for recording channels with partial-response equalization, as \textit{Partial-Response Neural Network} (PR-NN). We train bi-directional gated recurrent units (bi-GRUs)~\cite{bib::Schuster1997Birnn, bib::Chung2014Empirical} to recover the ISI channel inputs from noisy channel output sequences and evaluate the network performance when applied to continuous, streaming data. The computational complexity of PR-NN during the evaluation process is comparable to that of a Viterbi detector. The recording system on which the experiments were conducted uses a rate-2/3, (1,7) runlength-limited (RLL) code~\cite{bib::Weather1991Encoder} with an E\textsuperscript{2}PR4 partial-response channel target. Experimental results with ideal PR signals show that the performance of PR-NN detection approaches that of Viterbi detection in additive white gaussian noise (AWGN). Moreover, the PR-NN detector outperforms Viterbi detection and achieves the performance of  \textit{Noise-Predictive Maximum Likelihood} (NPML) detection in additive colored noise (ACN) at different channel densities. A PR-NN detector trained with both AWGN and ACN maintains the performance observed under separate training. Similarly, when trained with ACN corresponding to two different channel densities, PR-NN maintains its performance at both densities. Experiments confirm that this robustness is consistent over a wide range of signal-to-noise ratios (SNRs). Finally, PR-NN displays robust performance when applied to a more realistic magnetic recording channel with MMSE-equalized Lorentzian signals.

\end{abstract}

\section{Introduction}
\label{sec::intro}

\subsection{Background on magnetic recording}

The read/write process in longitudinal magnetic recording is often modeled as a continuous-time, linear time-invariant (LTI) system with bipolar input waveforms taking values $-1$ and $+1$ on  time intervals of fixed length $T_{c}$. The step response often takes the form of a unimodal pulse with a finite pulsewidth at half the maximum amplitude (PW50) whose size relative to the channel input interval ($PW50/T_{c}$) roughly determines the extent of inter-symbol interference (ISI) arising from adjacent transitions in the input waveform. When the channel output signals are synchronously sampled at intervals of $T_{c}$, the sampled system is often modeled as a finite impulse-response discrete-time LTI system~\cite{bib::Vasic1967Book}.


In order to facilitate the recovery of the input signal, magnetic recording systems often use some form of partial-response (PR) equalization. The PR equalizer shapes the readback signal in such a way that only a finite number of values are observed at sample times. 
For a range of linear recording densities, the sampled Lorentzian output response of the  PR-equalized  longitudinal recording channel is well modeled by the family of extended PR ``class  4'' (PR4) channels, denoted by E\textsuperscript{N-1}PR4~\cite{bib::Thapar1987pr4}. The impulse response of the E\textsuperscript{N-1}PR4 channel can be represented by  $x_{N}(D)=(1-D)(1+D)^{N}$, where $D$ is the delay operator and $N$ is a positive integer. The channel can be treated as a linear filter with integer coefficients, whose outputs are generated by a linear finite-state machine, where the number of states is $2^{N+1}$. In practice, the selected PR step response is governed by the channel step response and the choice of $PW50/T_{c}$.  When sampled, the PR-equalized noisy magnetic recording channel is often modeled, to first order, as a linear finite-state machine with additive, correlated noise.   

The sampled PR-equalized magnetic recording channel resembles a digital communication channel, and suitable detection and coding methods from communication theory can be beneficially applied.  The finite-state structure of the ISI channel is amenable to trellis-based sequence detection methods such as Viterbi detection~\cite{bib::Viterbi1967Dec}, which is optimal if the additive noise is assumed to be white Gaussian noise (AWGN). The combination of  PR channel equalization and Viterbi detection is referred to as PRML, an acronym for ``partial-response (PR) equalization with maximum-likelihood (ML) sequence detection''~\cite{bib::Cideciyan1992PRML, bib::Karabed1996Signal}. Channels can also use maximum \mbox{a-posteriori} probability (MAP) symbol detection based upon, for example, the BCJR algorithm~\cite{bib::Bahl1974BCJR}, which is also optimal in AWGN. When combined with PR equalization, the resulting system is referred to as PRMAP. To account for noise correlation (colored noise) due to equalization, Noise-Predictive Maximum Likelihood (NPML) detectors embed a noise prediction/whitening process into the branch metric computation of a Viterbi detector~\cite{bib::Coker1998NPML}. In longitudinal recording systems, NPML detectors offer significant performance gains over PRML detectors~\cite{bib::Coker1998NPML, bib::Moision1998ConstrainedNPML}. 

Magnetic recording systems often use both an error-correcting code and a constrained code. The general purpose of a constrained code is to improve the performance of the system by matching the characteristics of the recorded signals to those of the channel~\cite{bib::Immink1998Recorder}. In the context of PRML-type systems, the constrained code is often used to improve timing recovery as well as to increase the distinguishability of the sampled output sequences. Several classes of such ``distance-enhancing codes'' have been proposed, such as runlength-limited (RLL) codes~\cite{bib::Moision1998Distance} and  matched spectral null (MSN) codes~\cite{bib::Karabed1991Match}. 

In a coded PRML-type system, equalization plays a crucial role in determining the performance of the system. Advanced detectors must take into account the noise correlation and signal misequalization effects due to equalization~\cite{bib::Zheng2016adaptive}. The channel parameters in a magnetic hard disk storage system can vary due to several factors, including variations in temperature, head flying-height,  and track-dependent rotational speed~\cite{bib::Wu2009Perpendicular}. These parameter variations need to be taken into consideration in the design of the PR equalizer and they can also influence the performance of the detector. 
In this paper, we explore the use of machine learning methods -- specifically recurrent neural networks (RNNs) -- to design robust detectors for magnetic recording systems. 

\subsection{Machine learning for coded communication}

In recent years, machine learning has demonstrated its effectiveness in a wide range of applications, specifically in the fields of computer vision and natural language processing. The huge success of machine learning in these areas has triggered the interest of researchers to apply deep learning (DL) methods and neural networks (NNs) to channel coding problems. Nachmani \textit{et al.} proposed \textit{Weighted Belief Propagation} (WBP) algorithm using deep neural networks (DNNs) to decode noisy linear codewords~\cite{bib::Nachmani2016learn}. This approach was further studied by Lian \textit{et al.} in the context of simple scaling models with reduced complexity~\cite{bib::Lian2019WBP}. In~\cite{bib::Satorras2020FGCNNBP}, Satorras and Welling considered a hybrid model that combines belief propagation with an extension of graph nerual networks to factor graphs (FG-GNNs). Kim \textit{et al.} presented the first end-to-end communication system for feedback channels designed using deep learning, with RNN models for encoding and decoding~\cite{bib::Kim2018Deep}. Jiang \textit{et al.} incorporated some aspects of an iterative turbo decoder into an RNN-based end-to-end machine learning architecture that provides robust, near-optimal recovery of noisy turbo codewords, without BCJR knowledge~\cite{bib::Jiang2019Turbo}. Shlezinger \textit{et al.} introduced ViterbiNet, a decoder that  incorporates DNNs into the Viterbi algorithm~\cite{bib::Shlezinger2019Viterbinet}. 

With all of these learned communication systems, the length of codewords is quite limited because the training complexity grows exponentially in the length~\cite{bib::Grubber2017deepchannel}. Farsad and Goldsmith addressed this problem by creating a sliding bidirectional RNN to process a longer signal stream~\cite{bib::Farsad2017SBRNN}. Bennatan \textit{et al.} decoded codewords of an arbitrary block length by extracting the syndrome of the hard decisions and the channel output reliabilities~\cite{bib::Bennatan2018learning}. Tandler \textit{et al.} described a training method that gradually introduces code sequences with an increasing number of ones to limit the complexity, and used it to recover long convolutional codewords~\cite{bib::Tandler2019Rnn}. Nevertheless, during the evaluation stage, these NN-based decoders are not well suited to handling continuous streaming data, which would typically be produced by convolutional encoders and ISI channels.  

\subsection{Our Contribution}


In this work, we propose a novel NN architecture for detection of input-constrained \mbox{PR-equalized} magnetic recording channels, which we refer to as \textit{partial-response neural network} (PR-NN). The PR-NN detector is designed for application to continuous, streaming channel outputs. The sequential processing properties of RNN cells~\cite{bib::Chung2014Empirical, bib::Cho2014GRUcom}, 
including gated recurrent units (GRUs)~\cite{bib::Cho2014GRU} and long short-term memory (LSTM)~\cite{bib::Hochreiter1997lstm}, are naturally suited to the time-dependent outputs from PR channels. The primary component of our PR-NN architecture is a bi-directional gated recurrent unit (bi-GRU). (We settled on a GRU-based architecture after initial experiments indicated superior performance compared to an LSTM network. These results are not included here.)  

We train and evaluate the PR-NN under various scenarios: ideal PR channel outputs with AWGN, ideal PR channel outputs with additive colored noise (ACN) generated by a minimum mean squared error (MMSE) equalizer for the Lorentzian channel, and MMSE-equalized Lorentzian outputs with corresponding equalized noise. By training the model under multiple scenarios, we show that a single PR-NN detector can be used as a substitute for multiple classical detectors. Moreover, training over a range of channel signal-to-noise ratios (SNRs) and channel densities allows the PR-NN to adapt to a variety of channel conditions. Special training and evaluation techniques make the PR-NN compatible with detection of continuous, streaming data, with no constraint on sequence length, thus overcoming a key limitation of previous NN-based decoding strategies. The continuous decoding relies on a \textit{sliding-window} evaluation process. We also show that the computational complexity of the PR-NN detector is comparable with that of Viterbi detection.



We conduct our experiments using the E\textsuperscript{2}PR4 channel model, which matches the characteristics of the Lorentzian channel for high recording densities. The binary system inputs are constrained by a rate-2/3, (1,7)-RLL constrained sliding-block decodable finite-state encoder~\cite{bib::Weather1991Encoder}. The constrained codewords are mapped into binary channel inputs by a non-return-to-zero-inverse (NRZI) precoder with system function represented by $c(D)=1/(1+D)$. The resulting sequence is modulated to bipolar form to represent the magnetization pattern corresponding to the recording channel input~\cite{bib::Immink1998Recorder}. With AWGN and a reduced-state Viterbi detector that reflects the input constraint, the system serving as our benchmark achieves a $2.2$dB coding gain over the uncoded E\textsuperscript{2}PR4 system~\cite{bib::Armstrong1992Channel}. 

The bit error rate (BER) performance of the PR-NN detector compares favorably to that of the classical detectors -- PRML, PRMAP and NPML -- in the scenarios where they are known to perform well. More importantly, the PR-NN detector exhibits a robustness not shared by the other detectors when it is jointly trained in multiple scenarios. In fact, under joint training, PR-NN essentially maintains the performance that is achieved with separate training. 
These results suggest that robust detection architectures like PR-NN may hold promise for application in practical recording systems. 

\subsection{Outline}
The remainder of this paper is organized as follows. In Section~\ref{sec::systemdet}, we formulate the system architecture for the magnetic recording channel in more detail and describe the digital implementation used in our performance simulations. In Section~\ref{sec::rnndet}, we give a comprehensive discussion of the PR-NN detector, including network architecture, dataset generation, training methodology, evaluation procedure, and computational complexity. In Section~\ref{sec::exp}, we apply the PR-NN detector and determine its performance under several scenarios: individually trained models for E\textsuperscript{2}PR4 channels with AWGN and ACN, equalized-Lorentzian channels with ACN, and various jointly trained scenarios. We then use the simulation results to assess the robustness of PR-NN detection.

\section{System Architecture and Detectors}
\label{sec::systemdet}

\subsection{System model}
\label{subsec::system}

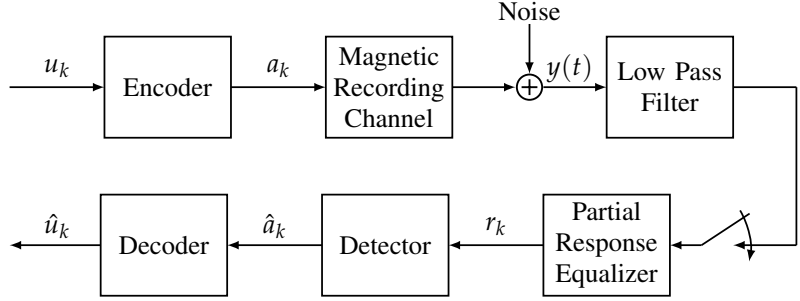
\begin{figure}[h]
	\begin{center}
		\resizebox{0.65\linewidth}{!}{
		\begin{tikzpicture}
		\draw [-latex, thick] (-1.5,0) -- (0,0);
		\node [align=center] at (-0.75,0.3) {$u_{k}$};
		\node [align=center] at (1,0) {Encoder};
		\draw [thick] (0,-0.8) rectangle (2,0.8);
		
		\draw [-latex, thick] (2,0) -- (3.5,0);
		\node [align=center] at (2.75,0.3) {$a_{k}$};
		\node [align=center] at (4.5,0) {Magnetic \\ Recording \\ Channel};
		\draw [thick] (3.5,-0.8) rectangle (5.5,0.8);
		\draw [-latex, thick] (5.5,0) -- (6.5,0);
		\node [align=center] at (7.35,0.3) {$y(t)$};
		
		\node [dspadder] at (6.72,0) {};
		\draw [-latex, thick] (6.72,1) -- (6.72,0.21);
		\node [align=center] at (6.72,1.2) {Noise};
		\draw [-latex, thick] (6.94,0) -- (7.94,0);
		\draw [thick] (7.94,-0.8) rectangle (9.94,0.8);
		\node [align=center] at (8.94,0) {Low Pass \\ Filter};
		
		\draw [thick] (9.94,0) -- (10.94,0);
		\draw [thick] (10.94,0) -- (10.94,-2.5);
		\draw [-latex, thick] (10.94,-2.5) -- (9.94,-2.5);
		\draw [-latex, thick] (9.44,-2.5) -- (8.94,-2.5);
		\draw [thick] (10.2,-2) -- (9.44,-2.5);
		\draw [-latex, thick] (10.05,-1.95) arc (30:-10:1.2);
		\node [align=center] at (7.94,-2.5) {Partial \\ Response \\ Equalizer};
		\draw [thick] (6.94,-3.3) rectangle (8.94,-1.7);
		\draw [-latex, thick] (6.94,-2.5) -- (5.44, -2.5);
		\node [align=center] at (6.19,-2.2) {$r_{k}$};
		
		\node [align=center] at (4.44,-2.5) {Detector};
		\draw [thick] (3.44,-3.3) rectangle (5.44,-1.7);
		\draw [-latex, thick] (3.44,-2.5) -- (1.94, -2.5);
		\node [align=center] at (2.69,-2.2) {$\hat{a}_{k}$};
		\node [align=center] at (0.94,-2.5) {Decoder};
		\draw [thick] (1.94,-3.3) rectangle (-0.06,-1.7);
		\draw [-latex, thick] (-0.06,-2.5) -- (-1.5, -2.5);
		\node [align=center] at (-0.75,-2.2) {$\hat{u}_{k}$};
		\end{tikzpicture}
	}
	\end{center}
	\vspace{-2ex}
	\caption{System architecture.}
	\label{fig::system}
	\vspace{-2ex}
\end{figure}

A block diagram of a magnetic-disk recording system is shown in Fig.~\ref{fig::system}. User data $\{u_{k}\}$ ($u_{k}\in\{0, 1\}$) is encoded using a $(d,k)$ run-length limited (RLL) code~\cite{bib::Immink1998Recorder}. The constrained codewords are then precoded and mapped into the symbol sequence $\{a_{k}\}$ ($a_{k}\in\{-1, +1\}$). The precoder maps a binary sequence to  the two-level channel input sequence. The precoding convention we use is nonreturn-to-zero-inverse (NRZI), where the precoder has system function $c(D)=1/(1+D)$. The modulation method is binary phase shift keying  (BPSK), where $0$ maps to $-1$ and $1$ maps to $+1$. 

During the write process, the two-level channel input is converted into a one-dimensional magnetization pattern on the magnetic medium in the disk. The disk spins with controlled speed, and the read and write heads effectively rotate over a track on the surface of the magnetic  medium~\cite{bib::Wu2009Perpendicular}. During the read process, the disk rotates beneath the read head, which senses the magnetic field induced by the magnetization pattern along the track. The read-back signal can be regarded as a linear superposition of the dipulse response corresponding to a positive pulse of width equal to a single channel bit duration at the input to the channel. Mathematically, the read-back signal $y(t)$ can be expressed as
\begin{equation}
y(t)=\sum_{i=-\infty}^{\infty}a_{i}q(t-iT_{c})+\eta(t)
\vspace{-1ex}
\end{equation}
where $T_{c}$ is the channel bit spacing, and sequence $\{a_{k}\}$ is written on the disk at a rate of $1/T_{c}$. The \textit{dipulse response} $q(t)$ is expressed as $q(t)=g(t)-g(t-T_{c})$, where $g(t)$ is the unit step response of the channel. The term $\eta(t)$ represents the additive white Gaussian noise process. 

The function $g(t)$ is often called the transition response of the recording system, and its characteristics are related to the specific design of the recording heads and magnetic medium. Recording systems are typically classified into two types: longitudinal~\cite{bib::Moision2000Magnetic} and perpendicular~\cite{bib::Wu2009Perpendicular}. The Lorentzian model for the transition response is commonly used for longitudinal recording systems. The $tanh$ function and error function approximation are widely used for perpendicular recording systems. In this work, we focus on longitudinal recording with Lorentzian transition response
\begin{equation}
g(t)=\frac{1}{1+(2t/PW_{50})^{2}}
\vspace{-1ex}
\end{equation}
where $PW_{50}$ is the single parameter of the Lorentzian model and denotes the pulsewidth at $50\%$ maximum amplitude. The recording density is characterized by the normalized density parameter $PW_{50}/T_{c}$.

The noisy channel output signal passes through a low pass filter (LPF). The filtered signal is sampled at the rate $1/T_{c}$, generating samples at times $t=kT_{c}$. The samples are filtered by a discrete-time equalizer which is designed to optimize detector performance. The most common scheme used for equalization and detection in longitudinal recording systems is partial-response maximum-likelihood detection (PRML). In this scheme, a \textit{finite impulse response} (FIR) equalizer is designed to equalize the channel response to a relatively short-duration partial-response (PR) target, and the channel input sequence is recovered from the equalized signal by  a maximum-likelihood  detector based on the Viterbi algorithm. The family of equalizers called ``Class-4'' and ``extended Class-4'' are often used in longitudinal magnetic recording, where the choice of equalizer target is matched to the channel density~\cite{bib::Thapar1987pr4}. The  general expression for the samples of the target equalized dipulse response, expressed as a $D$-transform polynomial, takes the form 
\begin{equation}
\begin{aligned}
x(D)&=(1-D)\alpha(D)=(1-D)(1+D)^{N} \\
&=x_{0}+x_{1}D+\cdots+x_{N+1}D^{N+1}
\end{aligned}
\vspace{-1ex}
\end{equation}
where $\alpha(D)=\alpha_{0}+\alpha_{1}D+\cdots+\alpha_{N}D^{N}$. When $N=1$, the channel is  called a Partial-Response Class-4 (PR4) channel. When $N\geq2$, the channels are call Extended Partial-Response Class-4, denoted individually as E\textsuperscript{N-1}PR4. 

There are many papers that address the design of the PR equalizer, such as~\cite{bib::Karabed1996Signal,bib::Moon1995MMSEequalizer}. A common design objective is the MMSE equalizer, which minimizes the mean squared error of the target PR signal and the equalized channel output. Since the channel parameters may vary across and even along tracks on a magnetic disk, the equalizer can be designed to be adaptive to the channel properties. Some adaptive equalization architectures for PR channels can be found in~\cite{bib::Vasic1967Book}. 

In the longitudinal recording system, if the target PR signal is chosen to be E\textsuperscript{N-1}PR4 signal, the coefficients of the PR equalizer can be optimized to achieve an overall transfer function that reflects the head/medium characteristics and the analog LPF frequency response. If we assume an ideal LPF, the equalizer coefficients $\{z_{i}\}$ can be specified as
\begin{equation}
\begin{aligned}
z_{i}&=\frac{1}{2\pi}\int_{-\pi}^{\pi}\frac{x(e^{-j\omega})}{Q(\omega)}e^{j\omega i}d\omega \\
&=\frac{1}{\pi^{2}}\sum_{\ell=0}^{N}\alpha_{i}\frac{(-1)^{\ell}e^{\pi PW_{50}/2}\cos(i\pi)-PW_{50}/2}{(PW_{50}/2)^{2}+(i-\ell)^{2}}
\end{aligned}
\label{eq::prequalizer}
\vspace{-1ex}
\end{equation}
where $Q(\omega)$ is the frequency response of the Lorentzian channel and $T_{c}=1$~\cite{bib::Moision2000Magnetic}. 

Thus the output $r_{k}$ of the PR equalizer in Fig.~\ref{fig::system} consists of an ideal PR  signal plus an additive  distortion. Mathematically, the equalizer output $r_{k}$ can be written as
\begin{equation}
\begin{aligned}
r_{k}=\sum_{i=0}^{N+1}x_{i}a_{k-i}+n_{k}
\end{aligned}
\label{eq::rk}
\vspace{-1ex}
\end{equation}
where $\{x_{i}\}$ are the coefficients of the target E\textsuperscript{N-1}PR4 channel and  $\{n_{k}\}$ denotes the additive distortion.  The distortion $n_k$ can be decomposed  as
\begin{equation}
\begin{aligned}
n_{k}&=\sum_{i}z_{i}\eta_{k-i}+q_{k} \\
&=\sum_{i}z_{i}\eta_{k-i}+(\sum_{i}\tilde{p}_{i}a_{k-i}-\sum_{i=0}^{N+1}x_{i}a_{k-i})
\end{aligned}
\label{eq::noise}
\vspace{-1ex}
\end{equation}
where the first summation represents additive colored noise corresponding to the equalized samples of the low-pass filtered white noise and the expression in parentheses represents the misequalization error $q_{k}$. Here $\{\tilde{p}_{i}\}$ correspond to the convolution of the PR equalizer taps with the sampled channel dipulse response.

As discussed above, given a PR target, a trellis-based Viterbi detector~\cite{bib::Viterbi1967Dec} is used to detect the data sequence from the noisy channel output sequence $\mathbf{r}$. For the Viterbi detector, the branch metric calculation is based on the squared-Euclidean distance between the noisy channel output sample and the targeted PR channel output sample labeling the particular branch. The combination of PR equalization with Viterbi detection is called PRML. The BCJR detector~\cite{bib::Bahl1974BCJR}, which is based upon a MAP symbol detection algorithm, is also of interest for data recovery. For the BCJR detector, the \textit{log-likelihood ratios} (LLRs) can be derived from forward recursion, backward recursion, and branch transition probability computations. The system combining PR equalization with BCJR detection is called PRMAP.
 
As shown in~(\ref{eq::noise}), the noise in the longitudinal recording system model is composed of colored noise and misequalization error, which cannot be modeled as AWGN. Therefore, the Viterbi detector is not an optimal sequence detector. NPML detection~\cite{bib::Coker1998NPML} combines a linear noise prediction/whitening filter with Viterbi detection. The coefficients of the noise predictor are designed to minimize the mean squared error (MSE) of the noise and the predictor output. Algorithmic details of these detection methods will be formulated in Section~\ref{subsec::classicdet}.

The decoder that recovers user data estimates $\hat{u}_{k}$  from channel input estimates $\hat{a}_{k}$ is implemented by means of a sliding-block decoder (which is here assumed to incorporate a $1+D$ post-coder operation). The decoder has memory $m$ and anticipation $a$, meaning that the current detected codeword, the previous $m$ detected codewords,  and the following $a$ detected codewords are all used to  determine the corresponding current user data word~\cite{bib::Marcus2001Book}.  

\subsection{Digital channel implementation}
\label{subsec::dig}

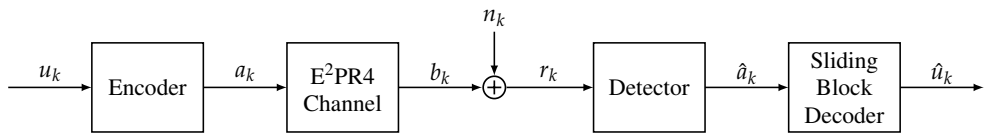
\begin{figure*}[h]
	\begin{center}
		\resizebox{0.8\linewidth}{!}{
		\begin{tikzpicture}
		\draw [-latex, thick] (-1.5,0) -- (0,0);
		\node [align=center] at (-0.75,0.25) {$u_{k}$};
		\node [align=center] at (1,0) {Encoder};
		\draw [thick] (0,-0.8) rectangle (2,0.8);
		\draw [-latex, thick] (2,0) -- (3.5,0);
		\node [align=center] at (2.75,0.25) {$a_{k}$};
		\node [align=center] at (4.5,0) {E\textsuperscript{2}PR4 \\ Channel};
		\draw [thick] (3.5,-0.8) rectangle (5.5,0.8);
		\draw [-latex, thick] (5.5,0) -- (7,0);
		\node [align=center] at (6.25,0.25) {$b_{k}$};
		\node [dspadder] at (7.22,0) {};
		\draw [-latex, thick] (7.22,1) -- (7.22,0.21);
		\node [align=center] at (7.22,1.25) {$n_{k}$};
		\draw [-latex, thick] (7.44,0) -- (9,0);
		\node [align=center] at (8.20,0.25) {$r_{k}$};
		\draw [thick] (9,-0.8) rectangle (11,0.8);
		\node [align=center] at (10,0) {Detector};
		\draw [-latex, thick] (11,0) -- (12.5,0);
		\node [align=center] at (11.75,0.25) {$\hat{a}_{k}$};
		\draw [thick] (12.5,-0.8) rectangle (14.5,0.8);
		\node [align=center] at (13.5,0) {Sliding \\ Block \\ Decoder};
		\draw [-latex, thick] (14.5,0) -- (16,0);
		\node [align=center] at (15.25,0.25) {$\hat{u}_{k}$};
		\end{tikzpicture}
	}
	\end{center}
	\vspace{-2ex}
	\caption{Discrete model for simulation, where $n_{k}$ represents the additive distortion.}
	\label{fig::systemsim}
	\vspace{-1ex}
\end{figure*}

In the following, we formulate the digital implementation of the magnetic recording system, as shown in Fig.~\ref{fig::system}, that we will use for our simulations. The outputs of the PR equalizer can be modeled as the outputs of a binary-input, linear ISI channel with additive distortion as derived in~(\ref{eq::rk}) and~(\ref{eq::noise}). In particular, we can model the ISI channel resulting from the magnetic recording channel, the LPF, the sampler and the PR equalizer in Fig.~\ref{fig::system} as an E\textsuperscript{N-1}PR4 channel, as shown in Fig.~\ref{fig::systemsim}. For our experiments, we focus on the specific case of $N=3$, namely E\textsuperscript{2}PR4, as this was widely used in practice. The outputs of the channel model can then be represented as
\begin{equation}
\begin{aligned}
r_{k}=b_{k}+\eta_{k}=\sum_{i=0}^{4}x_{i}a_{k-i}+n_{k}
\end{aligned}
\label{eq::isi}
\vspace{-1ex}
\end{equation}
where $x_{0}=1$, $x_{1}=2$, $x_{2}=0$, $x_{3}=-2$ and $x_{4}=-1$. Here $b_{k}$ denotes the noiseless output from the E\textsuperscript{2}PR4 channel and $a_{k}$ denotes the channel input. 

We now describe the finite-state channel representation of the input-constrained E\textsuperscript{2}PR4 channel. In order to improve the performance of PRML-type systems, several classes of codes with distance-enhancing properties have been proposed, such as forbidden list codes~\cite{bib::Immink1998Recorder} and \textit{matched spectral null} (MSN) codes~\cite{bib::Karabed1991Match}.  In~\cite{bib::Armstrong1992Channel}, Behrens and Armstrong show that $(1,\infty)$-RLL constrained codes provide a coding gain when applied to the E\textsuperscript{2}PR4 channel. To see this, note that a sequence satisfies the $(1,\infty)$-RLL constraint if the runs of $0$s between successive $1$s have length at least $1$~\cite{bib::Marcus2001Book}. In other words, consecutive $1$s are forbidden. This means that, in the precoded $(1,\infty)$-RLL sequence, the strings $101$ and $010$ are prohibited. The minimum squared-Euclidean distance between channel  output sequences corresponding to a closed error event (paths in the detector trellis that agree except on a finite number of branches) is $6$. These events correspond to the channel inputs $+1\ -1\ +1$ and $-1\ +1\ -1$. On the other hand, for the $(1,\infty)$ input-constrained channel, with these channel inputs forbidden, the minimum distance associated with a closed error event increases to $10$. (A more detailed discussion of this sort of distance analysis is found in~\cite{bib::Karabed1991Match}.) This offers the possibility of an effective $2.2$ dB gain in signal-to-noise ratio (SNR), ignoring the rate loss associated with the use of the constrained code, provided that the detector trellis is modified to reflect the constraints. Note that this gain also applies to the (1,7)-RLL input-constrained channel.  

\begin{figure}[h]
	\begin{center}
		\includegraphics[width=.5\linewidth]{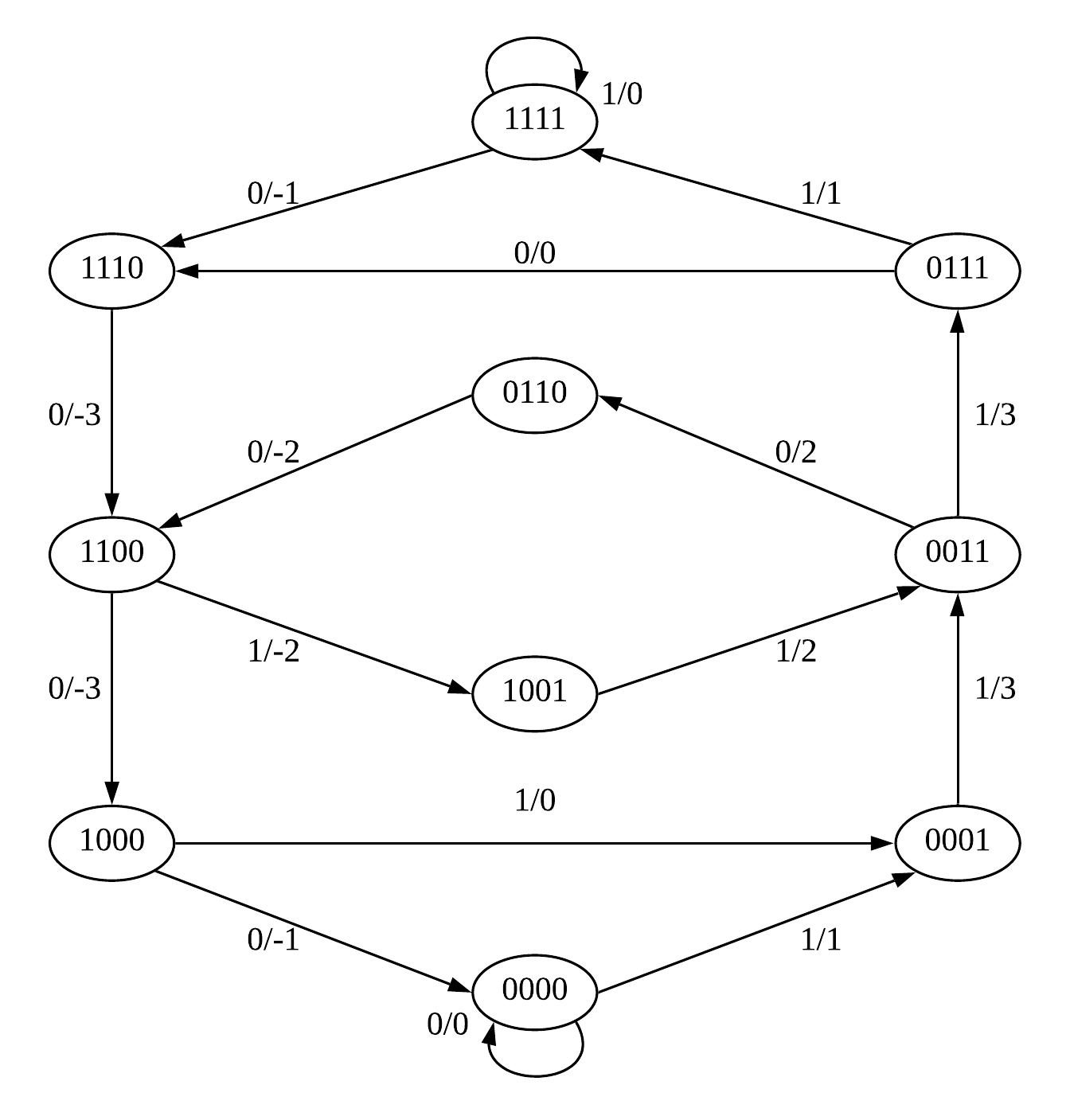}
	\end{center}
	\vspace{-3ex}
	\caption{$(1,\infty)$-RLL input-constrained E\textsuperscript{2}PR4 channel state machine.}
	\label{fig::statemachine}
\end{figure}

Incorporating the $(1,\infty)$-RLL constraint into the Viterbi detector for the E\textsuperscript{2}PR4 channel not only eliminates the dominant error events, but also reduces the required number of states in the detector trellis (i.e., the number of states realized by the channel finite state machine) from 16 to 10. To see this, for convenience, we ignore the BPSK modulation used to generate the bipolar channel inputs and, by a slight abuse of notation, we let  $a_{k}\in\{0, 1\}$ denote the channel inputs. At time $k$, channel state transitions from state  $\mathbf{s}_{k-1}=(a_{k-4}a_{k-3}a_{k-2}a_{k-1})$ to state  $\mathbf{s}_{k}=(a_{k-3}a_{k-2}a_{k-1}a_{k})$ with associated output $b_{k}$. This is represented in the channel state machine diagram by an edge from state $\mathbf{s}_{k-1}$ to state $\mathbf{s}_{k}$ with input/output label $a_k/b_k$. When the $(1,\infty)$-RLL constraint is applied, states $(0010)$, $(0100)$, $(0101)$, $(1010)$, $(1011)$ and $(1101)$ are eliminated, along with all their incoming and outgoing branches because they represent violations of the constraint. The resulting state machine diagram for the input-constrained E\textsuperscript{2}PR4 channel is shown in Fig.~\ref{fig::statemachine}. This reduced state machine provides the structure of the trellis that can be used at each time step of the reduced-state Viterbi detector. 

We selected the (1,7)-RLL constraint for our system since it is has been widely used in commercial magnetic tape and hard disk recording systems. For the encoder and decoder, we use the rate-2/3 Weathers-Wolf code~\cite{bib::Weather1991Encoder}, which achieves the minimum possible number of states for any rate-2/3 (1,7)-RLL code. The code is $(0,2)$-sliding-block decodable, meaning that the decoding algorithm can be implemented by a sliding-block decoder, where the current (length-3) codeword along with the following two codewords are used to determine the corresponding (length-2) input word. The encoder and decoder structures are given in~\cite{bib::Weather1991Encoder}. In the sliding-block decoder, a single channel bit error can affect the decoding of up to $3$ input words, or $6$ user bits, so the error propagation is limited. 

\subsection{Signal Detection Methods}
\label{subsec::classicdet}

In the following, we review three classical signal detection methods for the magnetic recording system. Given the noisy sequence $\mathbf{r}$, the detector will output an estimate $\hat{\mathbf{a}}$ of the channel input sequence. 

\begin{enumerate}
\item Viterbi detection: When we incorporate the $(1,\infty)$-RLL constraint into the E\textsuperscript{2}PR4 state machine, the $10$-state graph determines the trellis structure for the Viterbi detector. The Viterbi detector maximizes the likelihood (conditional probability) $\text{Pr}(\mathbf{r}|\mathbf{a})$~\cite{bib::Viterbi1967Dec}. When the noise is AWGN, the branch metric is the squared Euclidean distance. Specifically, the branch metric at time $kT_{c}$ from state $\mathbf{s}_{j}$ to state $\mathbf{s}_{m}$ takes the form 
\vspace{-1ex}
\begin{equation}
\begin{aligned}
\lambda_{k}(\mathbf{s}_{j},\mathbf{s}_{m})=[r_{k}-(a_{k}(\mathbf{s}_{m})+\sum_{i=1}^{4}x_{i}a_{k-i}(\mathbf{s}_{j}))]^{2}
\end{aligned}
\vspace{-1ex}
\end{equation}
where $a_{k}(\mathbf{s}_{m})$, $a_{k-1}(\mathbf{s}_{j})$, $a_{k-2}(\mathbf{s}_{j})$, $a_{k-3}(\mathbf{s}_{j})$ and $a_{k-4}(\mathbf{s}_{j})$ are the BPSK input values determined by hypothesized state transition $\mathbf{s}_{j}\rightarrow \mathbf{s}_{m}$. 
\item BCJR detection: The BCJR detection algorithm maximizes the a-posteriori probability $\text{Pr}(\mathbf{a}|\mathbf{r})$. The complete derivation can be found in~\cite{bib::Bahl1974BCJR}. In order to reduce the computational complexity, we make use of a modified algorithm, called \textit{max-log-map} detection, that uses the approximation  $\text{ln}\sum_{j}e^{a_{j}}\approx\max_{j}a_{j}$.
\item NPML detection: As mentioned above, the additive distortion in a realistic longitudinal recording system cannot be considered to be simply AWGN. A better approximation takes into account the noise coloration introduced by the equalizer as well as the misequalization error. In the presence of such noise, the Viterbi detector will not provide optimal detection and the system performance will be degraded. NPML detection introduces a noise prediction process into the branch computation of the Viterbi detector that significantly improves the system performance~\cite{bib::Coker1998NPML}. An estimate of the current noise sample, $\hat{n}_{k}$ is formed from previous $N_{p}$ noise samples, and then subtracted from $r_{k}$. The coefficients of the $N_{p}$-tap noise predictor $\{p_{i}\}$ are chosen to minimize the mean squared error between the noise $n_{k}$ and the estimate $\hat{n}_{k}$, 
\vspace{-1ex}
\begin{equation}
\begin{aligned}
\mathbb{E}[|n_{k}-\hat{n}_{k}|^{2}]=\mathbb{E}[|n_{k}-\sum_{i=1}^{N_{p}}n_{k-i}p_{i}|^{2}],
\end{aligned}
\vspace{-1ex}
\end{equation}
where $n_{k}$ takes the form in~(\ref{eq::noise}). The derivation of the MMSE predictor coefficients can be found in \cite{bib::Coker1998NPML}. 

The implementation of the NPML detector requires the use of tentative decisions from the survivor path memory associated with each state of the Viterbi detector. Mathematically, the branch metric at time $kT_{c}$ from state $\mathbf{s}_{j}$ to state $\mathbf{s}_{m}$ takes the form 
\vspace{-1ex}
\begin{equation}
\begin{aligned}
\lambda_{k}(\mathbf{s}_{j},\mathbf{s}_{m})=[r_{k}-\sum_{i=1}^{N_{p}}(r_{k-i}-(\sum_{l=0}^{4}x_{i}\hat{a}_{k-i-l}(\mathbf{s}_{j})))p_{i}-(a_{k}(\mathbf{s}_{m})+\sum_{i=1}^{4}x_{i}a_{k-i}(\mathbf{s}_{j}))]^{2}
\end{aligned}
\vspace{-1ex}
\label{eq::npml}
\end{equation}
where the terms $\hat{a}_{k-i}(\mathbf{s}_{j})$, $\hat{a}_{k-i-1}(\mathbf{s}_{j})$, $\hat{a}_{k-i-2}(\mathbf{s}_{j})$, $\hat{a}_{k-i-3}(\mathbf{s}_{j})$ and $\hat{a}_{k-i-4}(\mathbf{s}_{j})$ represent past decisions taken from the survivor path history associated with state $\mathbf{s}_{j}$, and $a_{k}(\mathbf{s}_{m})$, $a_{k-1}(\mathbf{s}_{j})$, $a_{k-2}(\mathbf{s}_{j})$, $a_{k-3}(\mathbf{s}_{j})$ and $a_{k-4}(\mathbf{s}_{j})$ are determined by the hypothesized state transition $\mathbf{s}_{j}\rightarrow \mathbf{s}_{m}$.
\end{enumerate}

\subsection{Detector implementation details}
In practice,  Viterbi detectors can retain only a finite path memory and must make an output decision after some fixed delay whether or not all survivor paths have merged. A common practice is to determine the trellis state with the minimum survivor path metric, and then to trace back along the path to the initial branch, whose label is then used to generate the estimated input/output symbol (or word, in the case of a convolutional code). Various estimates for a suitable traceback length $L_{tb}$ for convolutional codes have been proposed, based upon random coding analysis~\cite{bib::Forney1974Max}, code sequence properties~\cite{bib::Hemmati1977Truncation},  and experimentation~\cite{bib::McEliece1989Truncation, bib::Moision2008Truncation}. A reasonable rule of thumb for a rate-$r$ code with memory $\nu$ is 
\begin{equation}
\begin{aligned}
L_{tb}\approx A\frac{\nu}{1-r}
\end{aligned}
\vspace{-1ex}
\end{equation}
where $A$ is between 2 and 3. For rate $r=1/2$ codes, this agrees with the often cited estimate $L_{tb} \approx A \nu$ with $A$ between 4 and 6. Similar methods have been used to estimate $L_{tb}$ for Viterbi detection of ISI channels, and a reasonable choice for the traceback length, which we use to guide our experiments, is $L_{tb}\approx 5\nu$.

\begin{figure}[]
	\begin{center}
		\includegraphics[width=0.8\textwidth]{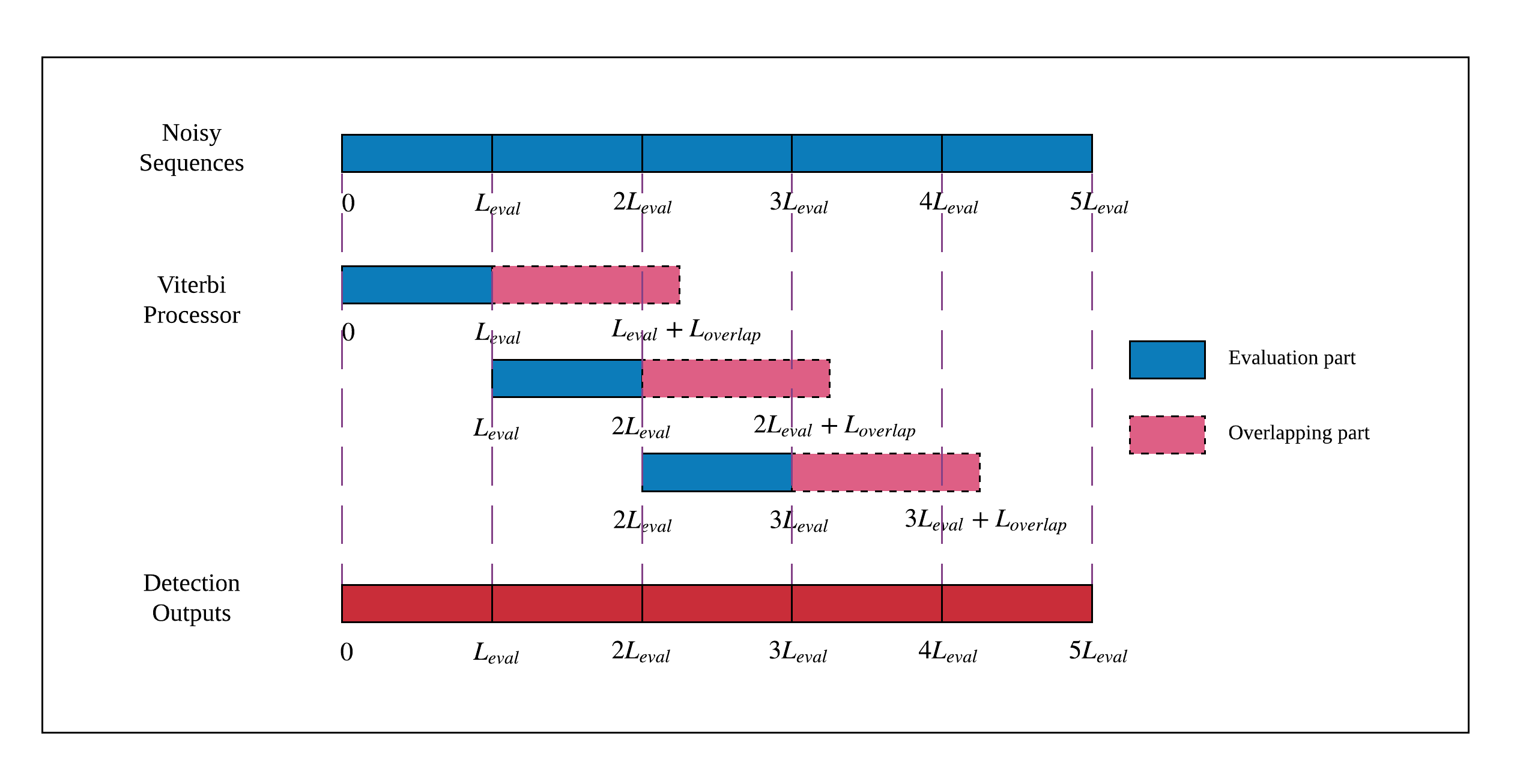}
	\end{center}
	\vspace{-4ex}
	\caption{Sliding-window evaluation process for Viterbi detection.}
	\label{fig::evalvit}
	\vspace{-2ex}
\end{figure}

Rather than decoding one symbol at each iteration of the survivor metric update procedure in the Viterbi detector, we will make use of a sliding-window approach. In the sliding-window decoder, successive blocks of a specified  ``evaluation length'' $L_{eval}$ are estimated, as illustrated schematically in Fig.~\ref{fig::evalvit}. The survivor metric computation starts at time $0$, and the survivor update procedure  is performed continuously as the first $L_{eval}+L_{overlap}$ received symbols arrive,
where $L_{overlap} \geq L_{tb}$. The detector then traces back along the survivor path corresponding to the state with the smallest survivor metric. The symbol estimates along the first $L_{eval}$ branches can be considered to be fairly reliable and the detector outputs these values. The first $L_{eval}$ branches of the survivor paths are then truncated, and the detector proceeds to extend the remaining portion of the survivor paths for another $L_{eval}$ steps, up to time $2L_{eval}+L_{overlap}$. The detector then traces back along the minimum metric survivor path and outputs the symbol estimates along the first $L_{eval}$ branches, corresponding to symbols at time $L_{eval}+1$ through $2L_{eval}$. This process is then repeated. In each successive block from time $mL_{eval}$ to time $(m+1)L_{eval}+L_{overlap}$, we refer to the first $L_{eval}$ steps as the evaluation part and the final $L_{overlap}$ steps as the overlap part.  The final $L_{overlap}$ symbols can be treated as dummy symbols, or a termination sequence of dummy symbols can be used to force  the detector to a known state, enhancing the reliability of the last group of estimated symbols.  We adopt the latter termination approach in our simulations.

The sliding-window approach is easily adapted to NPML detection. For the BCJR detector, a conceptually similar sliding-window approach can be  implemented using  a forward state metric processor and a pair of backward state metric processors \cite{bib::Viterbi1998Map}.

The length of the evaluation block, $L_{eval}$, can be chosen to be any size greater than or equal to one symbol, with the lower limit corresponding to conventional symbol-by-symbol Viterbi decoding. Larger sizes increase the required storage for survivor paths and the delay until the first symbol is decoded.  However, the use of longer survivor paths should increase the likelihood of survivor path merging, thereby improving reliability of decoding. In Section~\ref{sec::rnndet}, where we adopt a similar block streaming decoding architecture, the sizes of $L_{eval}$ and $L_{overlap}$ have an exponential effect on the size of the training dataset. This plays a role in the choice of these parameters. 
\vspace{-2ex}
\begin{remark}
	\label{rmk::tapnum}
	For the digital implementation of the longitudinal recording channel, we assume that the channel bit spacing $T_{c}=1$. The discrete channel model in Fig.~\ref{fig::systemsim} uses a $41$-tap model of the Lorentzian channel $\{g_{i}\}$ and a $21$-tap PR equalizer $\{z_{i}\}$. The NPML detector is implemented using   $4$-tap noise predictor,   $8$-tap noise predictor, and  $16$-tap noise predictors. 
	In the sliding-window evaluation process, the evaluation length $L_{eval}$ is $10$ and the overlapping length $L_{overlap}$ is $20$. 
	\qed
	\vspace{-5ex}
\end{remark}

\section{PR-NN: RNN-based Detection}
\label{sec::rnndet}

In this section, we present PR-NN (partial response - neural network), an RNN-based detection method for coded partial-response channels. We discuss the details of network architecture, dataset generation, training methodology, evaluation procedure, and computational complexity.

The main idea of PR-NN is to replace the classical detectors with a robust RNN-based detector. The motivation of our approach comes from the GRU-based decoder for noisy convolutional codewords in~\cite{bib::Tandler2019Rnn}. Although our paper focuses on the application of PR-NN to the  coded E\textsuperscript{2}PR4 channel for longitudinal magnetic recording channels, we believe the proposed approach can be easily adapted to other practical magnetic recording systems. 

\subsection{Neural Network Architecture}

Thanks to the rapid development of deep learning, many neural network architectures have emerged and shown their power in a variety of application domains. RNNs, in particular, have been adopted in several scenarios involving time-sequential data, making an RNN-based architecture a natural candidate for processing signals produced by a magnetic recording channel with inter-symbol interference and possibly correlated additive noise. 

In practice, we exploit more sophisticated recurrent hidden units that implement a gating mechanism~\cite{bib::Chung2014Empirical, bib::Cho2014GRUcom}, such as long short-term memory (LSTM) units~\cite{bib::Hochreiter1997lstm} and gated recurrent units (GRUs)~\cite{bib::Cho2014GRU}. Comparable performance has been found in networks using GRU and LSTM~\cite{bib::Chung2014Empirical}. Our aim in this work is to explore the potential of RNN-based signal detection in channels with ISI, rather than to compare the performance of different RNN units, so we will only consider networks based on GRU cells. 

\begin{figure}[t]
	\centering
	\begin{subfigure}[b]{0.45\columnwidth}
		\includegraphics[width=\linewidth]{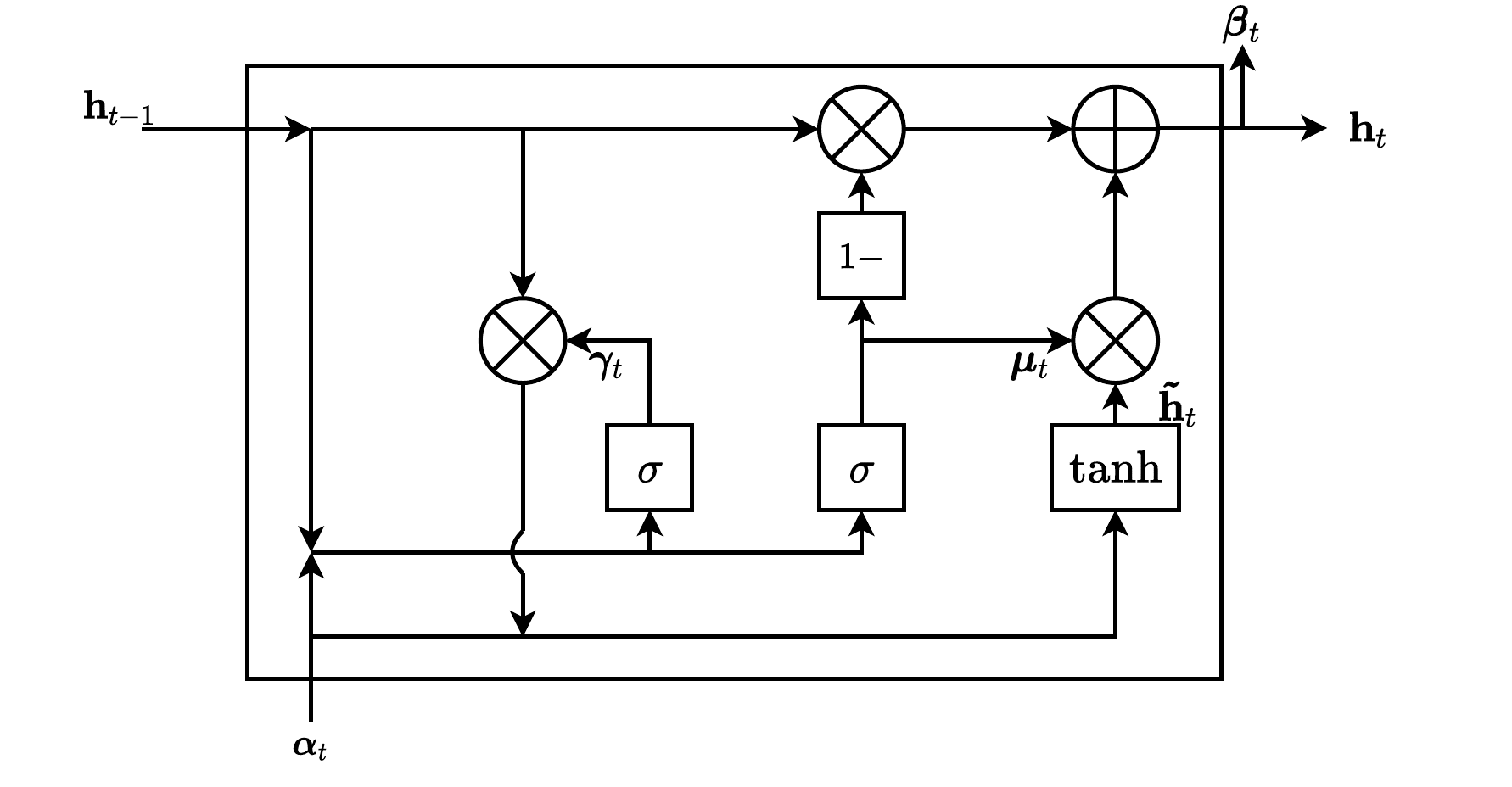}
		\caption{Structure of one GRU cell.} 
		\label{fig::grucell}
		\vspace{3ex}
	\end{subfigure}
	\begin{subfigure}[b]{0.45\columnwidth}
		\includegraphics[width=\linewidth]{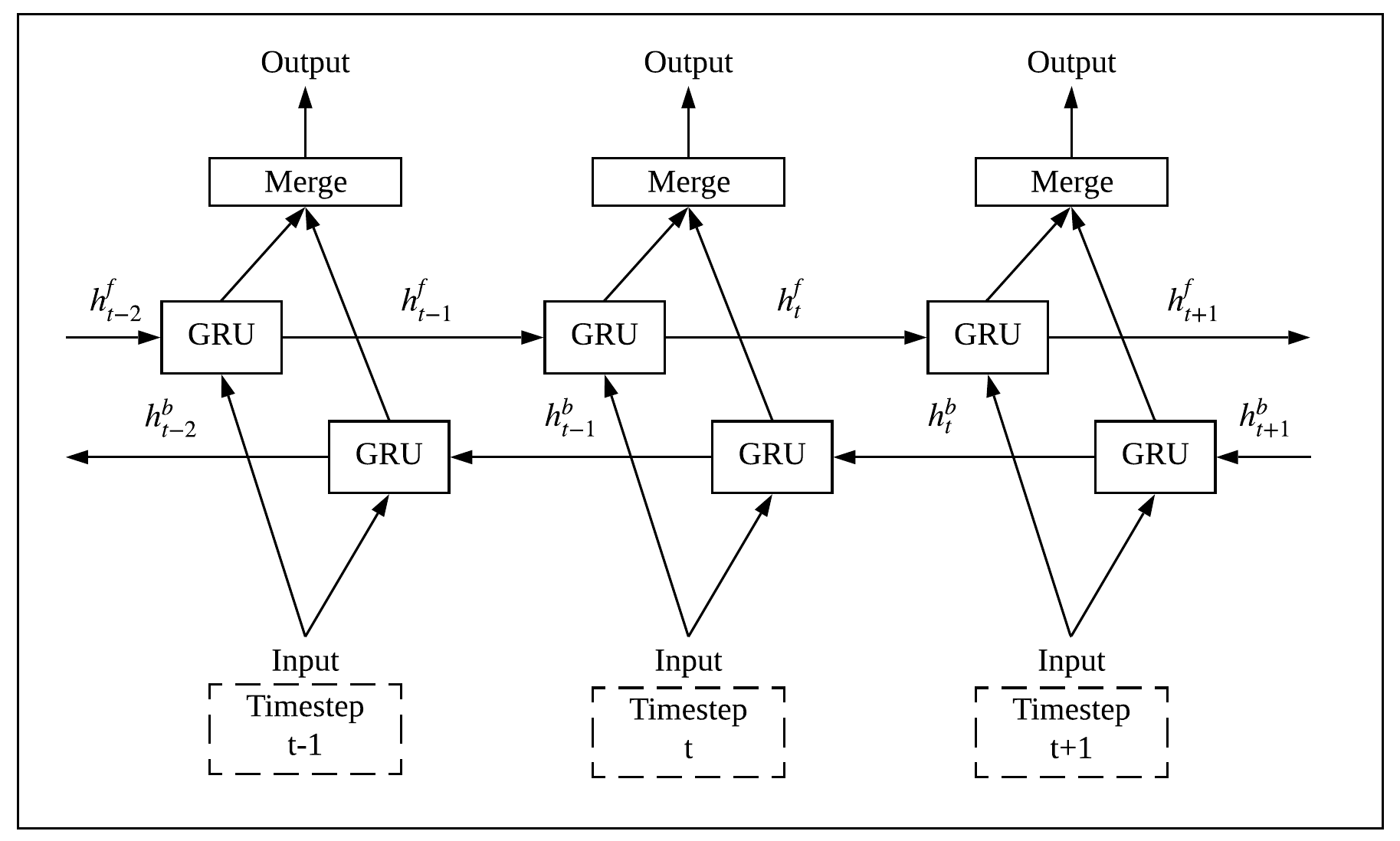}
		\caption{Structure of unfolded bi-directional GRUs.} 
		\label{fig::birnn}
		\vspace{3ex}
	\end{subfigure}
	\begin{subfigure}[b]{0.45\columnwidth}
		\includegraphics[width=\linewidth]{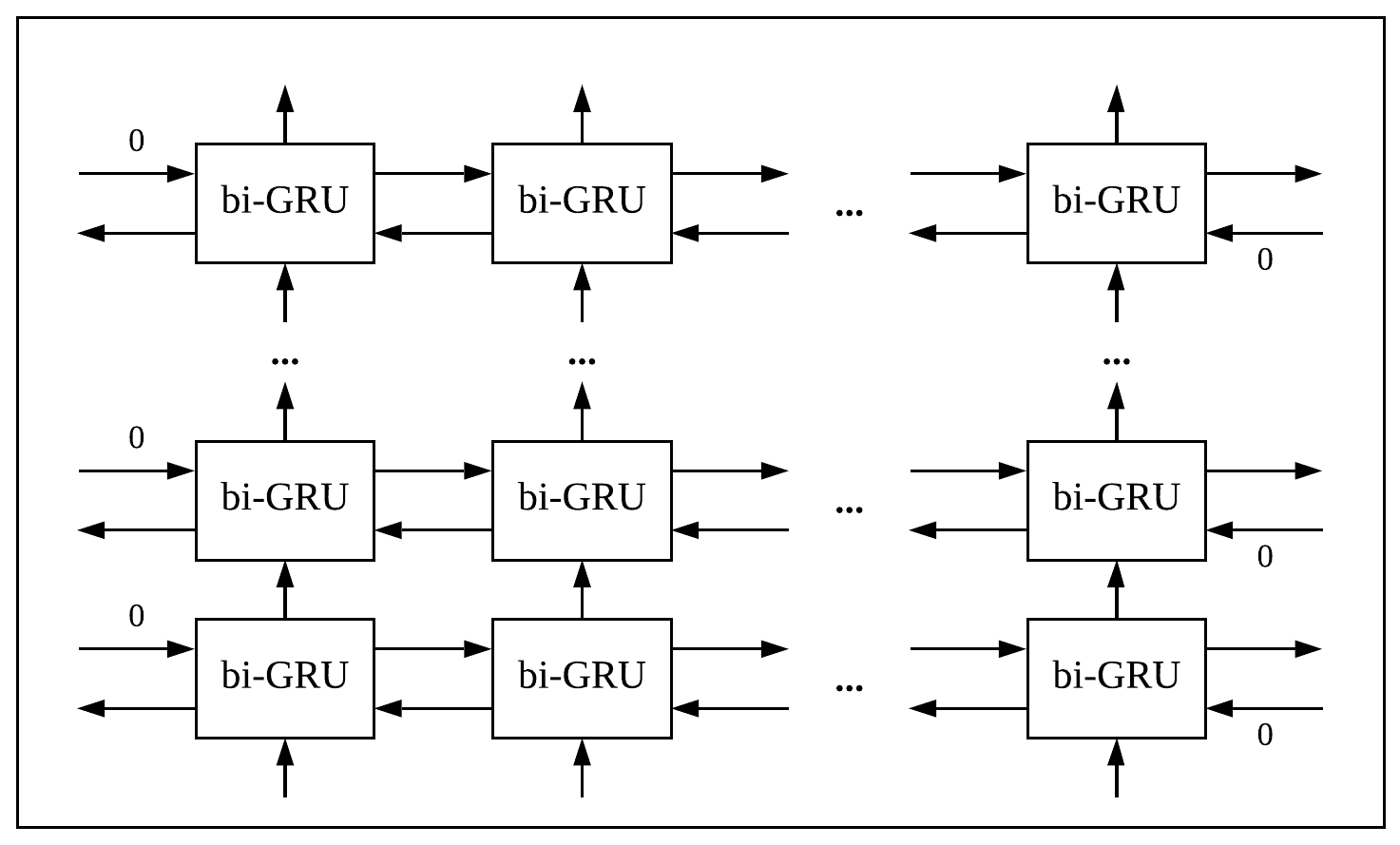}
		\caption{Model of multi-layer bi-directional GRUs.} 
		\label{fig::multibigru}
	\end{subfigure}
	\vspace{1ex}
	\caption{Model of bi-directional GRU network.}
	\label{fig::binet}
	\vspace{-3ex}
\end{figure}

A GRU schematic is shown in Fig.~\ref{fig::grucell}. The calculations in a GRU cell can be formulated as follows
\begin{equation}
\begin{aligned}
\boldsymbol{\gamma}_{t}&=\sigma(\mathbf{W}_{\alpha\gamma}\cdot\boldsymbol{\alpha}_{t}+\mathbf{b}_{\alpha\gamma}+\mathbf{W}_{h\gamma}\cdot\mathbf{h}_{t-1}+\mathbf{b}_{h\gamma})\\
\boldsymbol{\mu}_{t}&=\sigma(\mathbf{W}_{\alpha\mu}\cdot\boldsymbol{\alpha}_{t}+\mathbf{b}_{\alpha\mu}+\mathbf{W}_{h\mu}\cdot\mathbf{h}_{t-1}+\mathbf{b}_{h\mu}) \\
\mathbf{\tilde{h}}_{t}&=\tanh(\mathbf{W}_{\alpha\tilde{h}}\cdot\boldsymbol{\alpha}_{t}+\mathbf{b}_{\alpha\tilde{h}}+\boldsymbol{\gamma}_{t}\odot(\mathbf{W}_{h\tilde{h}}\cdot\mathbf{h}_{t-1}+\mathbf{b}_{h\tilde{h}})) \\
\mathbf{h}_{t}&=(\boldsymbol{1}-\boldsymbol{\mu}_{t})\odot\mathbf{\tilde{h}}_{t}+\boldsymbol{\mu}_{t}\odot\mathbf{h}_{t-1}
\end{aligned}
\label{eq::grucell}
\vspace{-1ex}
\end{equation}
where  we use standard notation for weight matrices $\mathbf W$, biases $\mathbf{b}$, and activation function $\sigma$.
In the calculations, the input at time step $t$ is $\boldsymbol{\alpha}_{t}\in\mathbb{R}^{m_{in}}$
, representing $m_{in}$ features.  
The hidden state at time step $t-1$ is $\mathbf{h}_{t-1}\in\mathbb{R}^{m_{h}}$.
The output at time step $t$ is $\boldsymbol{\beta}_{t}\in\mathbb{R}^{m_{out}}$.
The hidden state at time step $t$ is $\mathbf{h}_{t}\in\mathbb{R}^{m_{h}}$.
For a GRU cell, the hidden state $\mathbf{h}_{t}$ is the same as the output $\boldsymbol{\beta}_{t}$. 
The reset, update, and  new gates are represented by 
$\boldsymbol{\gamma}_{t}\in\mathbb{R}^{m_{h}}$, $\boldsymbol{\mu}_{t}\in\mathbb{R}^{m_{h}}$, and $\mathbf{\tilde{h}}_{t}\in\mathbb{R}^{m_{h}}$, 
respectively.
The Hadamard product is denoted by $\odot$. 

We adopt a bi-directional GRU (bi-GRU) architecture~\cite{bib::Schuster1997Birnn}, which can be understood as two separate GRU networks, one operating in the forward direction and the other operating in the backward direction. As illustrated in Fig.~\ref{fig::birnn}, in the forward (resp. backward) direction, the forward (resp. backward) GRU component of the bi-GRU at time step $t$ takes the hidden state $\mathbf{h}_{t-1}^{f}$ from time step $t-1$ (resp. the hidden state $\mathbf{h}_{t+1}^{b}$ from time step $t+1$) and produces the hidden state $\mathbf{h}_{t}^{f}$ for the next GRU cell at time step $t+1$ (resp. the hidden state $\mathbf{h}_{t}^{b}$ for the next GRU cell at time step $t-1$). The forward and backward outputs of the bi-GRU cells are concatenated at each time step. 
\vspace{-2ex}
\begin{remark}
\label{rmk::bigruset}
Our implementation of the GRU will incorporate multi-layer bi-GRU cells as illustrated in Fig.~\ref{fig::multibigru}. We suppose that the total number of time steps in each layer is $T_{r}$. The bi-GRU cell operates with simultaneous forward and backward passes at each time step $t$ ($1\leq t\leq T_{r}$). The input of the $i$-th layer ($i\geq 2$) is the hidden state of the previous layer. (We do not dropout any features from the GRU layer outputs.)
We set the default initial hidden states of bi-GRU cells (in both directions)  to $\boldsymbol{0}$; specifically, $\mathbf{h}_{0}^{f}=\boldsymbol{0}$ and $\mathbf{h}_{T_{r}+1}^{b}=\boldsymbol{0}$.
\qed
\vspace{-5ex}
\end{remark}

Referring to the coded E\textsuperscript{2}PR4 state machine in Fig~\ref{fig::statemachine}, we see a conceptual similarity between the forward pass of the bi-GRU and the operation of the Viterbi detector, with [current state/input] and [next state/output] corresponding to [previous hidden state/input] and [next hidden state/output], respectively. Similarly, the forward/backward passes of the bi-GRU bear a conceptual resemblance to the foward/backward passes of the BCJR detector. 

In order to  design an RNN-based detector for coded PR channels, we will have to train the network with noisy channel output sequences. The number of coded channel output sequences grows exponentially in the length of the channel input (approximately $2^{Rn}$, where $n$ is the length of the user input sequence and $R$ is the constrained code rate). This suggests the use of a block-oriented network architecture, with a limited block size. In order for the RNN-based detector to process continuous streaming channel outputs, we adopt a sliding-window approach, similar to the sliding-window implementations of the Viterbi detector and BCJR detector presented in Section~\ref{subsec::classicdet}, both of which process overlapping blocks. 

Referring to Fig.~\ref{fig::evalvit}, the idea is for the bi-GRU network to serve as the detection module for each block of length $L_{eval}$, using as the network input a length-($L_{eval} + L_{overlap}$) block of the recording channel output. However, when we feed a sequence into the multi-layer bi-GRU cells, the sequence needs to respect the default network initialization conditions in Remark \ref{rmk::bigruset}, i.e., the initial hidden state for the forward direction and the backward direction should be $\boldsymbol{0}$. We  assume that the default initial hidden state $\boldsymbol{0}$ corresponds to the state $(0000)$ in the coded E\textsuperscript{2}PR4 state machine. This poses a problem, because there is no guarantee that the block of inputs to the network correspond to a state sequence in the PR channel state machine that starts and ends in state $(0000)$. To compensate for this, we propose a \textit{zero compensation} approach, in which we append suitable starting and ending dummy values before and after each block to force sequences to start and end at state $(0000)$. The exact rules of the zero compensation approach are provided  in the next subsection. 

The input sequence to the bi-GRUs is therefore composed of four parts: starting dummy values, evaluation part, overlapping part, and ending dummy values. The respective lengths of these parts are denoted  $L_{start}$, $L_{eval}$, $L_{overlap}$, and $L_{end}$. The resulting total number of time steps in the bi-GRUs is $T_{r}=L_{start}+L_{eval}+L_{overlap}+L_{end}$. 

Now we specify the network components of the PR-NN detector. There are three kinds of layers: dense layers $\mathcal{D}(\mathbf{x})$, multi-layer bi-GRU cells $\mathcal{R}(\mathbf{x}, \mathbf{H}_{t}^{f}, \mathbf{H}_{t}^{b})$, and the sigmoid layer $\mathcal{S}(\mathbf{x})$. In this network, GRU cells are the key components. For time step $t$ ($1\leq t\leq T_{r}$), the details of three kinds of layers are listed below.
\begin{enumerate}
\item Dense layer $\mathcal{D}(\mathbf{x})$: given an input vector $\mathbf{x}\in\mathbb{R}^{m_{din}}$,  
a dense layer defines the following operation
\begin{equation}
\begin{aligned}
\mathbf{y}=\mathcal{D}(\mathbf{x})=\mathbf{W}_{d}\cdot\mathbf{x}+\mathbf{b}_{d}
\end{aligned}
\vspace{-1ex}
\end{equation}
where $\mathbf{y}\in\mathbb{R}^{m_{dout}}$ is the output of the dense layer.
\item Multi-layer bi-GRU cells $\mathcal{R}(\mathbf{x}, \mathbf{H}_{t}^{f}, \mathbf{H}_{t}^{b})$: The number of layers is denoted as $N_{r}$. The input vector 
is $\mathbf{x}\in\mathbb{R}^{m_{rin}}$, the forward (resp. backward) hidden state set is $\mathbf{H}_{t}^{f}=\{\mathbf{h}_{t}^{f1},\mathbf{h}_{t}^{f2},\cdots,\mathbf{h}_{t}^{fN_{r}}\}$ (resp. $\mathbf{H}_{t}^{b}=\{\mathbf{h}_{t}^{b1},\mathbf{h}_{t}^{b2},\cdots,\mathbf{h}_{t}^{bN_{r}}\}$), where $\mathbf{h}^*_{t}\in\mathbb{R}^{m_{rh}}$ is the hidden state vector 
and $t$ corresponds to the current time step. The output vector at time step $t$ is $\mathbf{y}\in\mathbb{R}^{m_{rout}}$.
The mathematical expression for the network operation is
\begin{equation}
\begin{aligned}
\mathbf{y}=\mathcal{R}(\mathbf{x}, \mathbf{H}_{t}^{f}, \mathbf{H}_{t}^{b})
\end{aligned}
\vspace{-1ex}
\end{equation}
(The calculations of the GRU cell were presented in~(\ref{eq::grucell}) and the structure of the multi-layer bi-GRU cells was formulated in Remark \ref{rmk::bigruset}.)
\item Sigmoid layer $\mathcal{S}(\mathbf{x})$: given an input vector $\mathbf{x}\in\mathbb{R}^{m_{sin}}$, the output $\mathbf{y}\in\mathbb{R}^{m_{sout}}$ of the sigmoid layer is 
\begin{equation}
\begin{aligned}
\mathbf{y}=\mathcal{S}(\mathbf{x}):y_{i}=\frac{1}{1+e^{x_{i}}}.
\end{aligned}
\vspace{-1ex}
\end{equation}
\end{enumerate}

The PR-NN contains the dense layers (Dense1 layer $\mathcal{D}_{1}(\mathbf{x})$ and Dense2 layer $\mathcal{D}_{2}(\mathbf{x})$), the multi-layer bi-GRU cells ($\mathcal{R}(\mathbf{x}, \mathbf{H}_{t}^{f}, \mathbf{H}_{t}^{b}, t)$), and the Sigmoid layer ($\mathcal{S}(\mathbf{x})$). The input to the PR-NN is derived from a length-$T_{r}$ noisy channel output sequence, $\mathbf{r}=\{r_{1},r_{2},\cdots,r_{T_{r}}\}$, where the total number of time steps in PR-NN is also $T_{r}$. 
To reflect  the memory of the E\textsuperscript{2}PR4 channel, we transform the sequence $\mathbf{r}$ into  the network  input vector $\underline{\mathbf{r}}'=\{\mathbf{r}^{(1)},\mathbf{r}^{(2)},\cdots,\mathbf{r}^{(T_{r})}\}$, where $\mathbf{r}^{(k)}=\{r_{k-4},r_{k-3},r_{k-2},r_{k-1},r_{k}\}$. 

\begin{figure}[]
	\begin{center}
		\includegraphics[width=.50\linewidth]{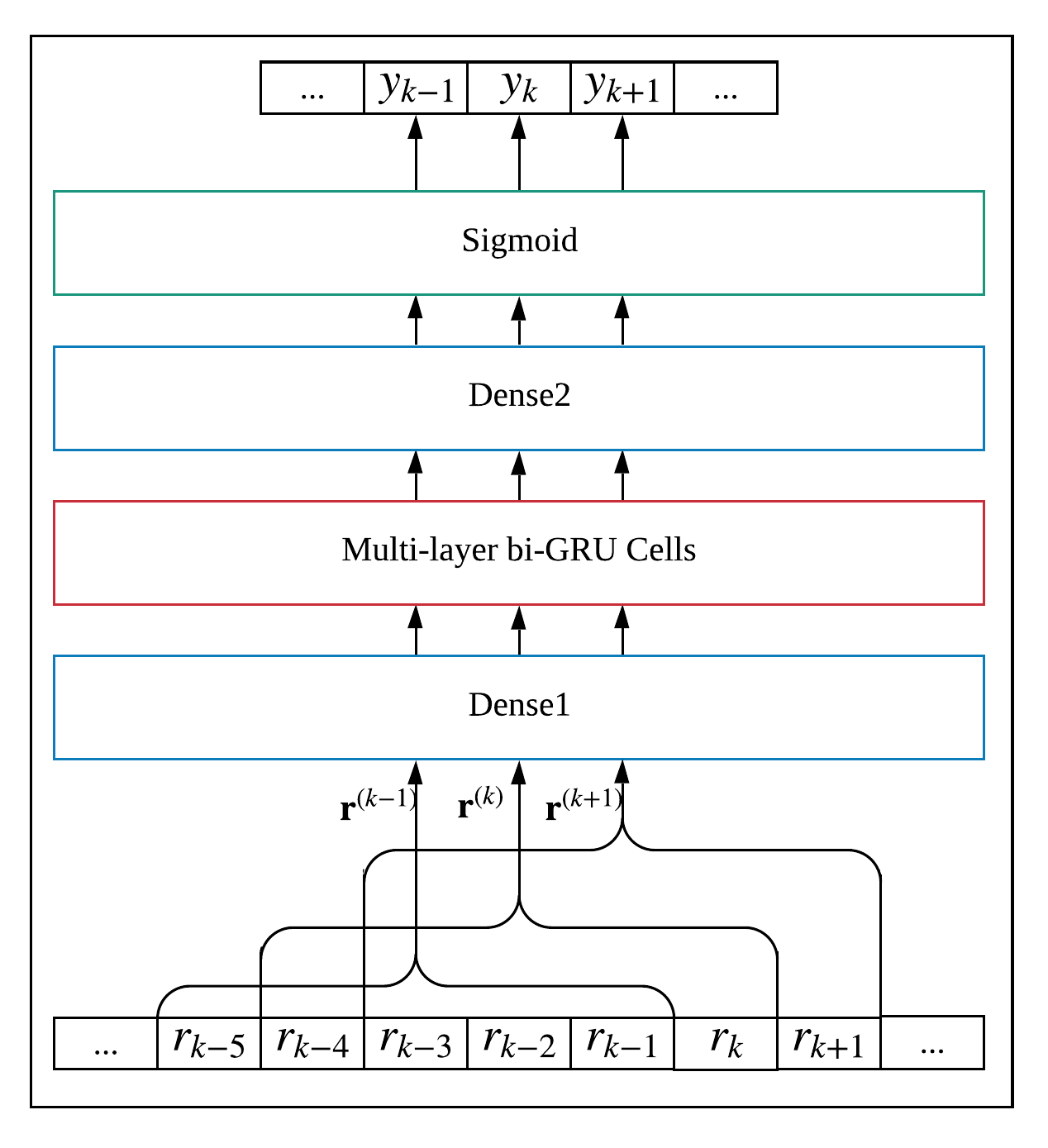}
	\end{center}
	\vspace{-3ex}
	\caption{Network architecture for proposed RNN-based detection.}
	\label{fig::netarch}
	\vspace{-2ex}
\end{figure}

The network architecture is shown in Fig.~\ref{fig::netarch}. Taking into account the starting and ending dummy values, we discard the the outputs of multi-layer bi-GRUs for the time steps $1\leq k\leq L_{start}$ and $T_{r}-L_{end}+1\leq k\leq T_{r}$.
The output vector is $\mathbf{y}=\{y_{L_{start}+1},y_{L_{start}+2},\cdots,y_{T_{r}-L_{end}}\}\in\mathbb{R}^{(L_{start}+L_{eval})}$ where, for time step $L_{start}+1\leq k\leq T_{r}-L_{end}$, the output is given by
\begin{equation}
\begin{aligned}
y_{k}=\mathcal{S}(\mathcal{D}_{2}(\mathcal{R}(\mathcal{D}_{1}(\mathbf{r}^{(k)}), \mathbf{H}_{k}^{f}, \mathbf{H}_{k}^{b}))).
\end{aligned}
\vspace{-1ex}
\end{equation}
Here we use time index $k$, rather than $t$, to be consistent with the indexing in the symbol sequences generated by the magnetic recording channel.

%

\vspace{-2ex}
\begin{remark}
\label{rmk::network}
In our experiments, for the network architecture, we chose $L_{start}=L_{end}=5$, $L_{eval}=10$. The length of the overlapping part is $L_{overlap}=20$, which is the same as the truncation depth in the Viterbi detector. Thereby, the total number of time steps in PR-NN is $T_{r}=40$. The parameters $L_{eval}$ and $L_{overlap}$ are also used in the simulations for the classical detection methods. For Dense1 layer, the number of input features is $5$ and the number of output features is $5$. For the multi-layer bi-GRU cells, the number of layers is $N_{r}=4$ and the number of features in a hidden state is $50$. For Dense2 layer, the number of input features is $100$ and the number of output features is $1$. 
\qed
\vspace{-4ex}
\end{remark}

\subsection{Data Acquisition}

The PR-NN detector recovers PR channel inputs from  noisy PR channel outputs.  The training dataset of noisy channel outputs is created as follows.  First, the coded  E\textsuperscript{2}PR4 channel state  machine in Fig.~\ref{fig::statemachine} is used to generate  a length-($L_{eval}+L_{overlap}$) channel input sequence from an arbitrarily chosen initial state,  and then suitable distortion is added. In our experiments, we consider three kinds of noise generated from the longitudinal recording system: AWGN, ACN generated by the MMSE PR equalizer for the Lorentzian channel, and total  distortion noise $n_{k}$ generated according to~(\ref{eq::noise}). In order to train the PR-NN to adapt to different SNRs,  the noise in the training set also reflects a range of SNRs. 

Then, we use a \textit{zero compensation} rule to ensure the network inputs satisfy the initial settings of bi-GRU cells, which require the corresponding starting and ending states of the PR channel  state  machine to be $(0000)$.  
A string of $L_{start}=5$ starting dummy values  is  prepended  to the noisy channel output sequence according to the initial  state, as indicated in Table~\ref{table::dummy}. As will be described in Section~\ref{subsec::eval}, PR-NN uses a sliding-window evaluation process, in which the starting state  for each truncated input block is determined by the symbols in the previously recovered evaluation block. The starting  dummy values are then determined by the zero  compensation rule, and since the starting state  is  assumed  to be correct, we do not add noise to the starting dummy values. The final state  of the path used to generate the input sequence determines a path to state $(0000)$ and a corresponding  string of  $L_{end}=5$ ending dummy values,  also shown in  Table~\ref{table::dummy}. However, since the ending state will be unknown during evalution, we add noise to these ending dummy values in the training sequence to represent noisy outputs corresponding to an unknown ending state  sequence.  
The training label for this training sequence is the length-($L_{eval}+L_{overlap}$) channel input sequence.

During evalution, length-($L_{eval}+L_{overlap}$) truncated blocks of a continuous streaming noisy channlel output sequence will be applied to the PR-NN detector. The evaluation labels are the corresponding detected PR-channel input sequences. In order to process the truncated blocks, a string of starting dummy values of length $L_{start}$ is prepended, using the starting state derived from the previously recovered evaluation block and Table~\ref{table::dummy}. An all-zero string of length $L_{end}$ is appended to the block, reflecting the fact that the  final state of the truncated block is unknown.

Example 1 illustrates the use of Table~\ref{table::dummy} in the generation of dummy values. 



\begin{table}[]
	\centering
	\scalebox{0.85}{
	\begin{tabular}{@{}ccc@{}}
		\hline
		State & Starting dummy values & Ending dummy values  \\ 
		\hline
		$(0000)$     		& $\{0, 0, 0, 0, 0\}$   		& $\{0, 0, 0, 0, 0\}$           \\
		$(0001)$     		& $\{0, 0, 0, 0, 1\}$   		& $\{3, 2, -2, -3, -1\}$        \\
		$(0011)$     		& $\{0, 0, 0, 1, 3\}$   		& $\{2, -2, -3, -1, 0\}$        \\
		$(0110)$     		& $\{0, 0, 1, 3, 2\}$        	& $\{-2, -3, -1, 0, 0\}$        \\
		$(0111)$     		& $\{0, 0, 1, 3, 3\}$        	& $\{0, -3, -3, -1, 0\}$        \\
		$(1000)$     		& $\{1, 3, 2, -2, -3\}$      	& $\{-1, 0, 0, 0, 0\}$          \\
		$(1001)$     		& $\{1, 3, 2, -2, -2\}$	    	& $\{2, 2, -2, -3, -1\}$        \\
		$(1100)$     		& $\{0, 1, 3, 2, -2\}$       	& $\{-3, -1, 0, 0, 0\}$         \\
		$(1110)$     		& $\{0, 1, 3, 3, 0\}$        	& $\{-3, -3, -1, 0, 0\}$        \\
		$(1111)$     		& $\{0, 1, 3, 3, 1\}$        	& $\{-1, -3, -3, -1, 0\}$       \\
		$\text{Unknown}$  	& $\{0, 0, 0, 0, 0\}$        	& $\{0, 0, 0, 0, 0\}$           \\
		\hline
	\end{tabular}}
	\caption{Starting and ending dummy values for each state in the coded E\textsuperscript{2}PR4 state machine. ``Unknown'' means unknown starting or ending state for the sequence.}
	\label{table::dummy}
	\vspace{-4ex}
\end{table}
\vspace{-2ex}
\begin{example}
If the starting state is $(1001)$, a path which forces the sequence from $(0000)$ to $(1001)$ is 
$$(0000)\!\!\rightarrow\!\!(0001)\!\!\rightarrow\!\!(0011)\!\!\rightarrow\!\!(0110)\!\!\rightarrow\!\!(1100)\!\!\rightarrow\!\!(1001).$$ Thus the corresponding starting dummy values for state $(1001)$ are $\{1,3,2,-2,-2\}$. If the ending state is $(1001)$, a path which drives the sequence from $(1001)$ to $(0000)$ is $$(1001)\!\!\rightarrow\!\!(0011)\!\!\rightarrow\!\!(0110)\!\!\rightarrow\!\!(1100)\!\!\rightarrow\!\!(1000)\!\!\rightarrow\!\!(0000).$$ Thus the corresponding ending dummy values for state $(1001)$ are $\{2,2,-2,-3,-1\}$. During training, noise will then be added to these values. \qed
\vspace{-5ex}
\end{example}

%

\subsection{Training Methodology}

The training dataset is fed into the network and we compare the outputs from the network with the training labels. The comparison metric is the loss function. The loss function between an output vector $\mathbf{y}$ and its corresponding channel input $\mathbf{a}$ can be defined as 
\begin{equation}
\begin{aligned}
\mathcal{L}=\sum_{k=L_{start}+1}^{L_{start}+L_{overlap}}-a_{k}\cdot\text{log}(y_{k})-(1-a_{k})\cdot\text{log}(1-y_{k})
\end{aligned}
\vspace{-1ex}
\end{equation}

To address the complexity of training the nework with the entire channel output space,  we adopted the $\textit{a-priori ramp-up}$ training method in \cite{bib::Tandler2019Rnn}. 
Instead of beginning the training with independent,  uniform user data $u_{k}\sim\text{Bern}(0.5)$, we start training with user data $u_{k}\sim\text{Bern}(p)$ ($p<0.5$) and gradually increase $p$ to $0.5$.
In our case, the training data $r_{k}$ is generated from the precoded constrained symbols $a_{k}$, and $a_{k}$ is determined by the user data $u_{k}$. The biasing probability $p(\text{ep})$ in $u_{k}\sim\text{Bern}(p)$ is a function of the epoch number $\text{ep}$, defined as 
\begin{equation}
\begin{aligned}
p(\text{ep})=\begin{cases}
0.1+0.01\cdot\floor{\text{ep}/\#\text{step}}, & \text{ep}\leq 40\cdot\#\text{step} \\
0.5, & \text{ep}> 40\cdot\#\text{step}
\end{cases}
\end{aligned}
\vspace{-1ex}
\end{equation}
where $\#\text{step}$ is the number of epochs between increments in the probability.  After every $\#\text{step}$ epochs, the probability $p(\text{ep})$ will increase  by $0.01$ until $0.5$ is reached.

\subsection{Evaluation Process}
\label{subsec::eval}

The outputs $y_k$ of the network are real values in the range $[0,1]$. These are converted to binary detector outputs  $\hat{a}_{k}$ using an indicator function: $\hat{a}_{k}=\mathbbm{1}_{\{x>0.5\}}(y_{k})$.
%
%

\begin{figure}[]
	\begin{center}
		\includegraphics[width=.8\linewidth]{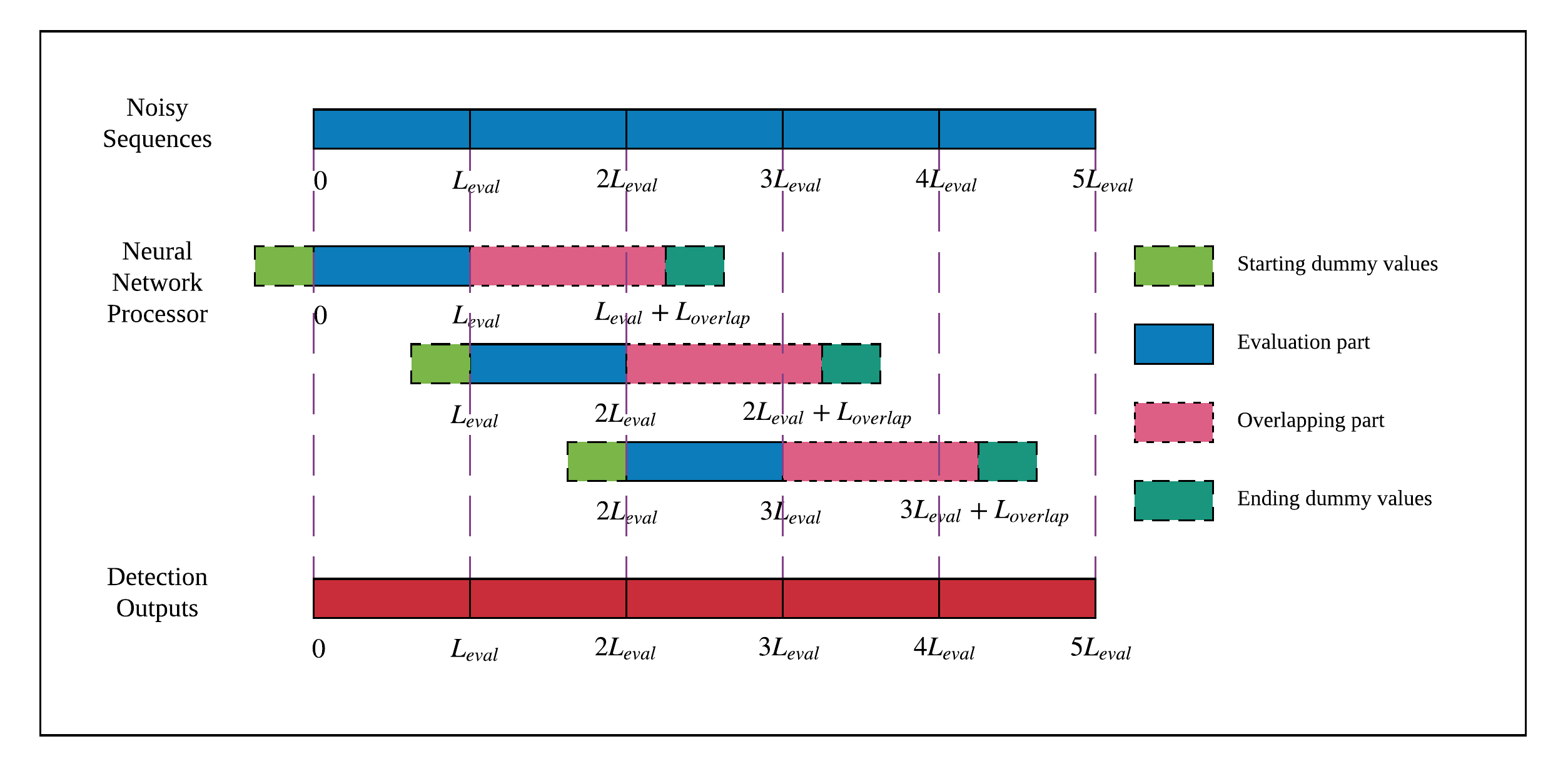}
	\end{center}
	\vspace{-4ex}
	\caption{Sliding-window evaluation process for PR-NN detector.}
	\label{fig::eval}
	\vspace{-2ex}
\end{figure}

The sliding-window concept used in the evaluation process is shown in Fig.~\ref{fig::eval}. 
We assume the state machine has initial state  $(0000)$.
When the first  block  of $L_{eval}+L_{overlap}$ symbols are received, the PR~-~{NN} detector prepends to the block the starting dummy values corresponding to state $(0000)$ and appends the ending dummy values corresponding to the unknown state. 
The resulting sequence is processed by the network, producing the detector outputs for the first length-$L_{eval}$ block. 
The last 4 bits of the recovered block determine the starting state  for the next detection stage.
When an additional $L_{eval}$ output symbols are received, the sliding window shifts by $L_{eval}$  positions and begins processing the next length-($L_{eval}+L_{overlap}$) truncated block. The starting dummy values for this block depend on the previously recovered starting state, and the ending dummy values correspond to the unknown ending state.  The resulting length-$L_{r}$ block is then processed by the network.  This procedure continues until the entire  stream of noisy channel outputs is processed.


For a noisy channel output sequence of length $L$, the evalulation metric is the bit error rate (BER) between the input $\mathbf{a}$ and the detection result $\hat{\mathbf{a}}$,  defined as 

\begin{equation}
\begin{aligned}
\text{BER}=\frac{1}{L}\sum_{k=1}^{L}\mathbbm{1}_{\hat{a}_{k}\neq a_{k}}(\hat{a}_{k})
\end{aligned}
\vspace{-1ex}
\end{equation}
%

\begin{remark}
\label{rmk::expsetting}
For the training set, the noise is generated for SNR values (in dB) in the set $\mathbb{S} =\{8.5, 9.0, 9.5, 10.0, 10.5\}$.
Throughout the training, Adam optimizer \cite{bib::Kingma2014adam} was used with learning rate $10^{-3}$. 
The value $\#\text{step}$ used in the  a-priori ramp-up training was set to $50$.
\qed
\vspace{-5ex}
\end{remark}

\subsection{Computational Complexity Analysis}
\label{subsec::complexity}

In this section, we compare the computational complexity of the Viterbi dectector and PR-NN detector.
We denote the length of noisy channel output sequence  by $L$ and the number of states in the input-constrained channel trellis by $N_{st}$.  Note that $N_{st}\leq 2^{v}$.


For a Viterbi detector, the \textit{add-compare-select} operation \cite{bib::Viterbi1967Dec} at each state requires augmenting  the survivor metrics for up to  $N_{st}$ paths, comparing the results, and selecting the minimum. The overall complexity can  be estimated as  $N_{st}^{2}L\mathcal{T}_{0}$, for an  appropriate  constant $\mathcal{T}_{0}$, or 
$O(N_{st}^2 L)=O(2^{2\nu} L)$. 

%

We  now consider the evaluation process of  PR-NN. Noting that the parameters  of  the dense layers and bi-GRU layers are independent of  the time step, we  let $\mathcal{T}_{NN}$ denote the computational complexity of one time step in the network. 
The number of multiplication operations and the number of addition operations in a combined matrix-vector product and vector addition operation such as 
$\mathbf{W}\cdot\boldsymbol{x}+\mathbf{b}$ are both equal to the number of elements in the matrix. 
We let $\mathcal{T}_{1}$  denote the complexity of a pair of multiplication and addition operations,  and assume  that the complexity of the activation and indicator  functions can be ignored.  Denoting the numbers of weights in Dense1 layer, multi-layer bi-GRU cells (one time step) and Dense2 layer as $n_{D1}$, $n_{G}$ and $n_{D2}$, respectively, we see that $\mathcal{T}_{NN} \leq (n_{D1}+n_{G}+n_{D2})\mathcal{T}_{1}$. 

%

The  processing of  each truncated block of length $L_{eval}+L_{overlap}$ requires $(L_{start}+L_{eval}+L_{overlap}+L_{end})$ steps,  where $L_{start}=L_{end}=L_{dummy}$ and  $L_{overlap}$ has the form $A\nu$, for some constants $L_{dummy}$ and  $A$. Thus,   the associated complexity is nominally $C=(2L_{dummy}+L_{eval}+A\nu)\mathcal{T}_{NN}$. Since the multi-layer bi-GRU outputs corresponding to dummy values can be ignored by the Dense2 layer, the complexity is actually 
$C'=C - 2L_{dummy}n_{D2}\mathcal{T}_{1}$. Processing the entire network  input stream involves approximately $L/L_{eval}$ blocks, so the overall complexity is approximately $LC'/L_{eval} = O(\nu L)$.  

Thus, the complexity of PR-NN compares favorably to that of Viterbi detection in PRML. Analogous complexity analysis for PRMAP and NPML detection leads to similar conclusions.


\section{Experimental Results}
\label{sec::exp}

In this section, we  present our experimental results for  PR-NN detection of the coded E\textsuperscript{2}PR4 channel. 
We consider three scenarios. First, we train the network separately on ideal PR signals with AWGN or ACN generated by a PR equalizer.  Then we use joint training on different combinations of AWGN and ACN, as well as on  a combination of ACN corresponding to different  channel  densities.  Finally, we train the network with ``realistic'' equalized Lorentzian channel  signals and distortions that include colored noise and misequalization errors.  Together, these experiments shed light on the robustness of PR-NN detection. 
We note that under each scenario, with the  a-priori  ramp-up approach, PR-NN training converges after $(40\cdot\#\text{step})$ epochs based  on monitoring the loss  function. 


\subsection{Experimental Setup}
\label{subsec::expsetup}

In the first   scenario,  we train PR-NN with only one kind of noise, i.e., AWGN or ACN. For ACN,  the colored noise is generated by applying the MMSE PR equalizer to AWGN samples. According to~(\ref{eq::prequalizer}), the ACN is affected by the channel density parameter $PW_{50}/T_{c}$, so we use two different,  representative values of $PW_{50}/T_{c}$ in our experiments. In the training process, the batch size for each SNR in $\mathbb{S}$ is $30$. 


In the second scenario, we assess PR-NN robustness by  training a single network to adapt to two different types of noise: (a) AWGN and ACN at  $PW_{50}/T_{c}=2.54$,  or (b) ACN at $PW_{50}/T_{c}=2.54$ and at $PW_{50}/T_{c}=2.88$.
The training batch size settings for case (a) are shown in lines 1, 2, and 3  and for case (b) in line 4 of Table~\ref{table::snrbtsize2}.


In the third scenario, the channel outputs represent a more  realistic, MMSE-equalized Lorentzian channel with misequalization errors and ACN.  To assess robustness to  different channel densities,  we  train the  PR-NN with separate  datasets at  $PW_{50}/T_{c}=2.54$ or $PW_{50}/T_{c}=2.88$, and then jointly with a dataset combining the two densities. Batch sizes are shown in lines 5, 6, and 7 of Table~\ref{table::snrbtsize2}.


\begin{table}[]
	\centering
	\scalebox{0.8}{
	\begin{tabular}{@{}c|ccccc|ccccc@{}}
		\hline 
		SNR & 8.5dB & 9.0dB & 9.5dB & 10.0dB & 10.5dB & 8.5dB & 9.0dB & 9.5dB & 10.0dB & 10.5dB \\
		\hline
		\hline
		Noise type & \multicolumn{5}{c|}{White noise} & \multicolumn{5}{c}{Colored noise ($PW_{50}/T_{c}=2.54$)}  \\
		\hline
		Experiment 2.1 & 30 & 30 & 30 & 30 & 30 & 30 & 30 & 30 & 30 & 30  \\
		Experiment 2.2 & 40 & 40 & 40 & 40 & 40 & 20 & 20 & 20 & 20 & 20   \\
		Experiment 2.3 & 50 & 50 & 50 & 50 & 50 & 10 & 10 & 10 & 10 & 10   \\
		\hline
		Noise type & \multicolumn{5}{c|}{Colored noise ($PW_{50}/T_{c}=2.54$)} & \multicolumn{5}{c}{Colored noise ($PW_{50}/T_{c}=2.88$)}  \\
		\hline
		Experiment 2.4 & 30 & 30 & 30 & 30 & 30 & 30 & 30 & 30 & 30 & 30  \\
		\hline
		Noise type & \multicolumn{5}{c|}{``Realistic'' system ($PW_{50}/T_{c}=2.54$)} & \multicolumn{5}{c}{``Realistic'' system ($PW_{50}/T_{c}=2.88$)}  \\
		\hline
		Experiment 3.1 & 30 & 30 & 30 & 30 & 30 & - & - & - & - & -  \\
		Experiment 3.2 & - & - & - & - & - & 30 & 30 & 30 & 30 & 30  \\
		Experiment 3.3 & 30 & 30 & 30 & 30 & 30 & 30 & 30 & 30 & 30 & 30  \\
		\hline
	\end{tabular}
	}
	\caption{Batch size settings for the training datasets and evaluation cases in each experiment  of the three scenarios.}
	\label{table::snrbtsize2}
\end{table}

\subsection{Scenario 1: Individual Training Experiments}

In scenario 1, PR-NN is only trained with one noise, i.e., AWGN, ACN ($PW_{50}/T_{c}=2.54$), ACN($PW_{50}/T_{c}=2.88$), where the batch size is $30$ for each SNR in $\mathbb{S}$. The evaluation results over the corresponding channels are described below.

\begin{figure}[]
	\begin{center}
		\includegraphics[width=0.45\columnwidth]{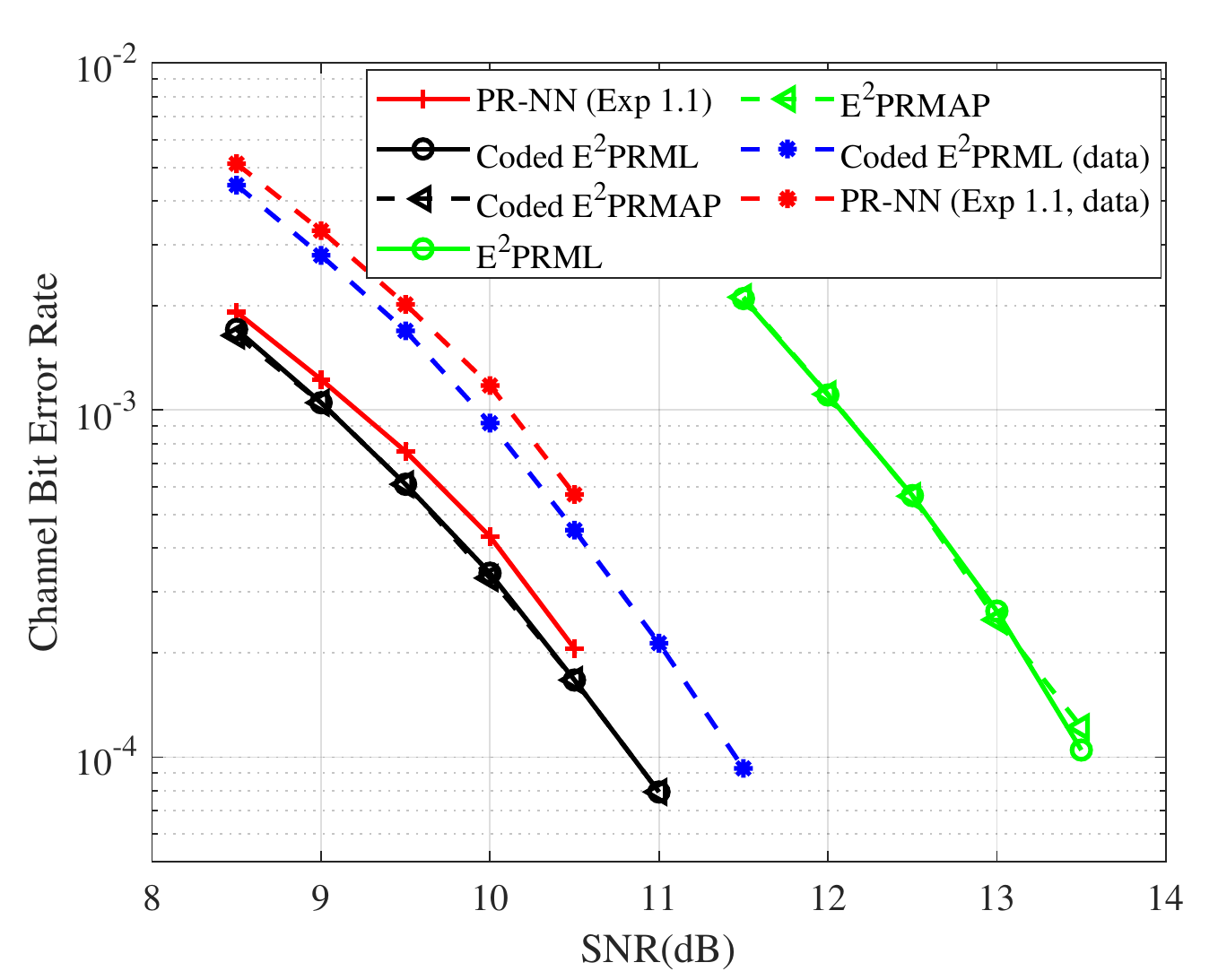}
	\end{center}
	\vspace{-3ex}
	\caption{Scenario 1: Individual training with AWGN.}
	\label{fig::exp1awgn}
	\vspace{-3ex}
\end{figure}

\textbf{Experiment 1.1}: We trained the PR-NN with AWGN. As shown in the BER plot in Fig.~\ref{fig::exp1awgn}, coded E\textsuperscript{2}PRML achieves the expected  2.2dB gain over uncoded E\textsuperscript{2}PRML with Viterbi detection~\cite{bib::Armstrong1992Channel}. In both cases, E\textsuperscript{2}PRMAP, implemented by \textit{max-log-map} approximation, performs essentially the same as E\textsuperscript{2}PRML. The user data BER of the coded E\textsuperscript{2}PRML channel  suffers a loss of about  0.9dB due to error propagation of the sliding-block decoder. 
 

We see that the PR-NN achieves performance within 0.1dB of the optimal Viterbi detector. There is a similar gap in performance for the user data BER. (In subsequent experiments, we present only the BER results at the detector output, not the user data results.)


At all SNRs, the histograms of error positions observed within an evaluation block of length $L_{eval}=10$ for the PR-NN detector and the Viterbi detector are approximately uniform, indicating that the overlapping part is providing ``reliable'' state information for the evaluation part.

\begin{figure}[t]
	\centering
	\begin{subfigure}[b]{0.45\columnwidth}
		\includegraphics[width=\linewidth]{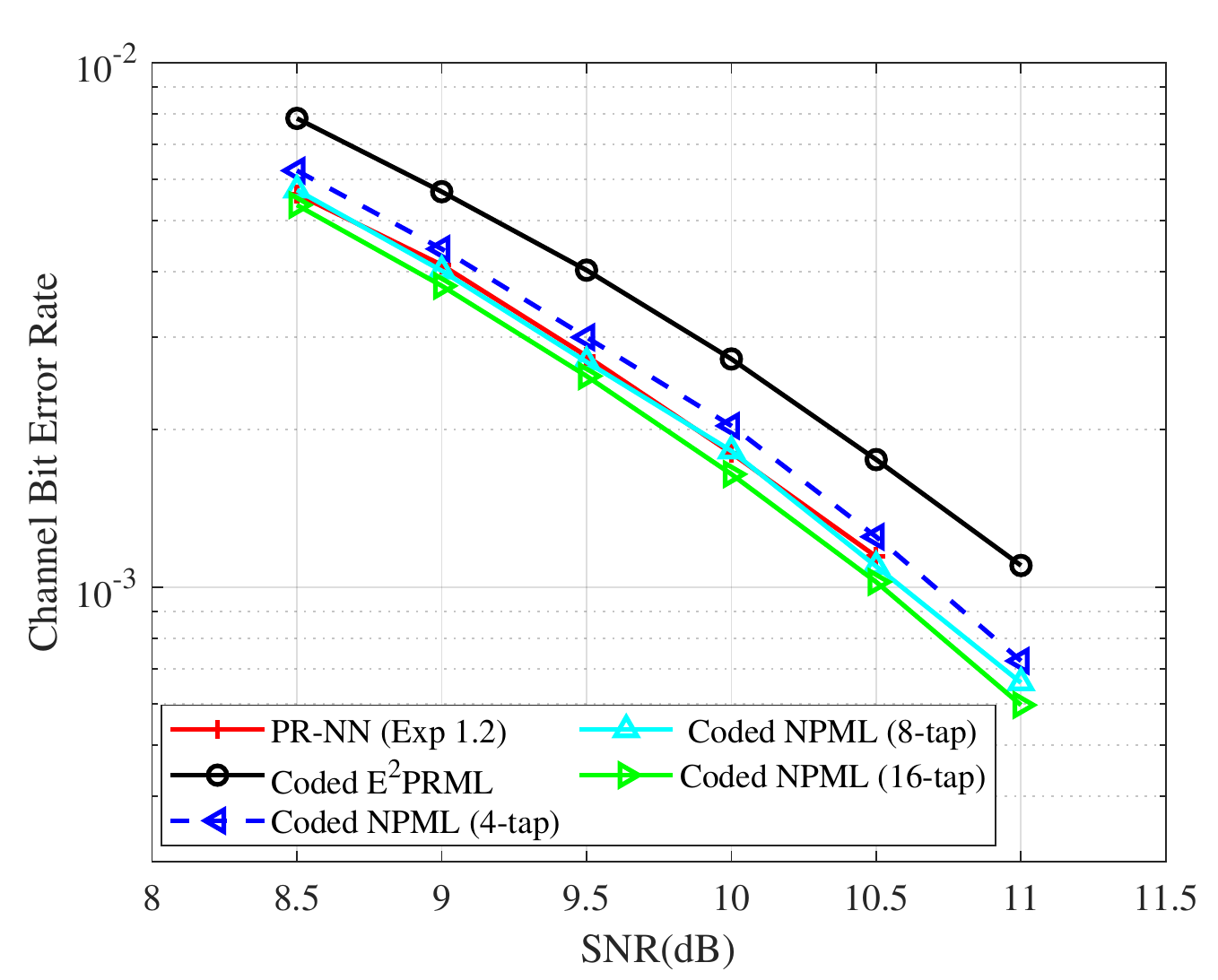}
		\caption{$PW_{50}/T_{c}=2.54$} 
		\label{fig::exp1acn254}
	\end{subfigure}
	\begin{subfigure}[b]{0.45\columnwidth}
		\includegraphics[width=\linewidth]{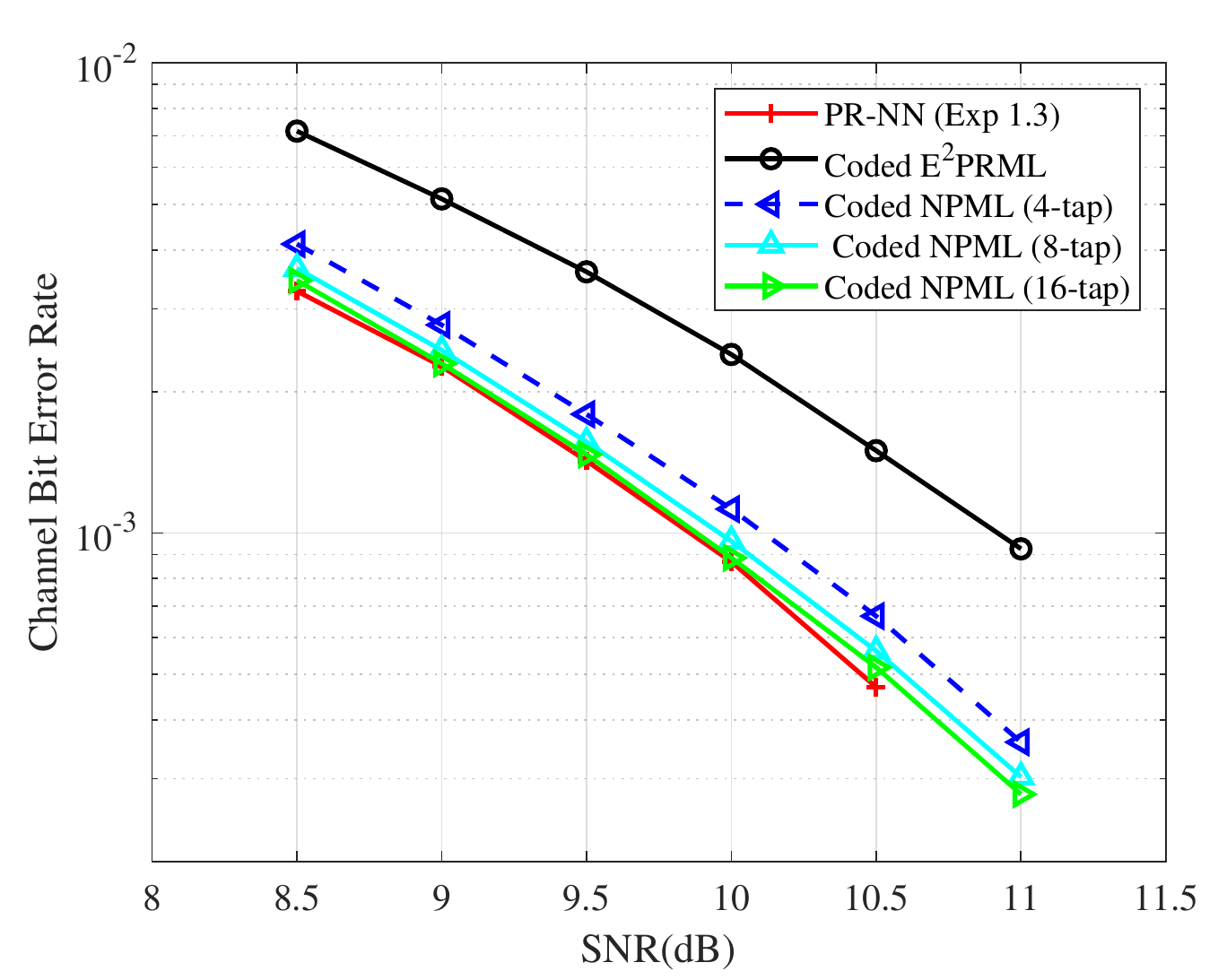}
		\caption{$PW_{50}/T_{c}=2.88$} 
		\label{fig::exp1acn288}
	\end{subfigure}
	\vspace{1ex}
	\caption{Scenario 1: Individual training  with ACN (for two channel densities).}
	\label{fig::exp1acn}
	\vspace{-1ex}
\end{figure}

\textbf{Experiment 1.2}: The PR-NN detector is trained and evaluated with ACN ($PW_{50}/T_{c}=2.54$). Referring to Fig.~\ref{fig::exp1acn254}, we see that the performance of coded E\textsuperscript{2}PRML is degraded in colored noise. The NPML detectors with $4$-tap, $8$-tap, and $16$-tap predictors realize gains of 0.4dB, 0.5dB, and 0.6dB gain, respectively, over   coded E\textsuperscript{2}PRML. Note that for these experiments, the NPML  detector designs assume  no misequalization error.   The PR-NN detector has very similar performance to the $8$-tap NPML detector.

%

\textbf{Experiment 1.3}: The PR-NN detector is trained and tested with ACN ($PW_{50}/T_{c}=2.88$). Referring to Fig.~\ref{fig::exp1acn288}, we see that, in this case,  the NPML detectors with   $4$-tap, $8$-tap, and $16$-tap predictors realize gains of 1.3dB, 1.5dB, and 1.55dB, respectively, over   coded E\textsuperscript{2}PRML. The PR-NN performance is very close to that of the $16$-tap NPML detector. In fact, at SNR=10.5dB, the BER achieved by PR-NN is even slightly better.


\subsection{Scenario 2: Joint Training Experiments}

In scenario 2, we train a single PR-NN using a dataset that combines noisy outputs representing different recording channels, and then evaluate its performance on both channels. Four different situations are considered.

\begin{figure}[t]
	\centering
	\begin{subfigure}[b]{0.45\columnwidth}
		\includegraphics[width=\linewidth]{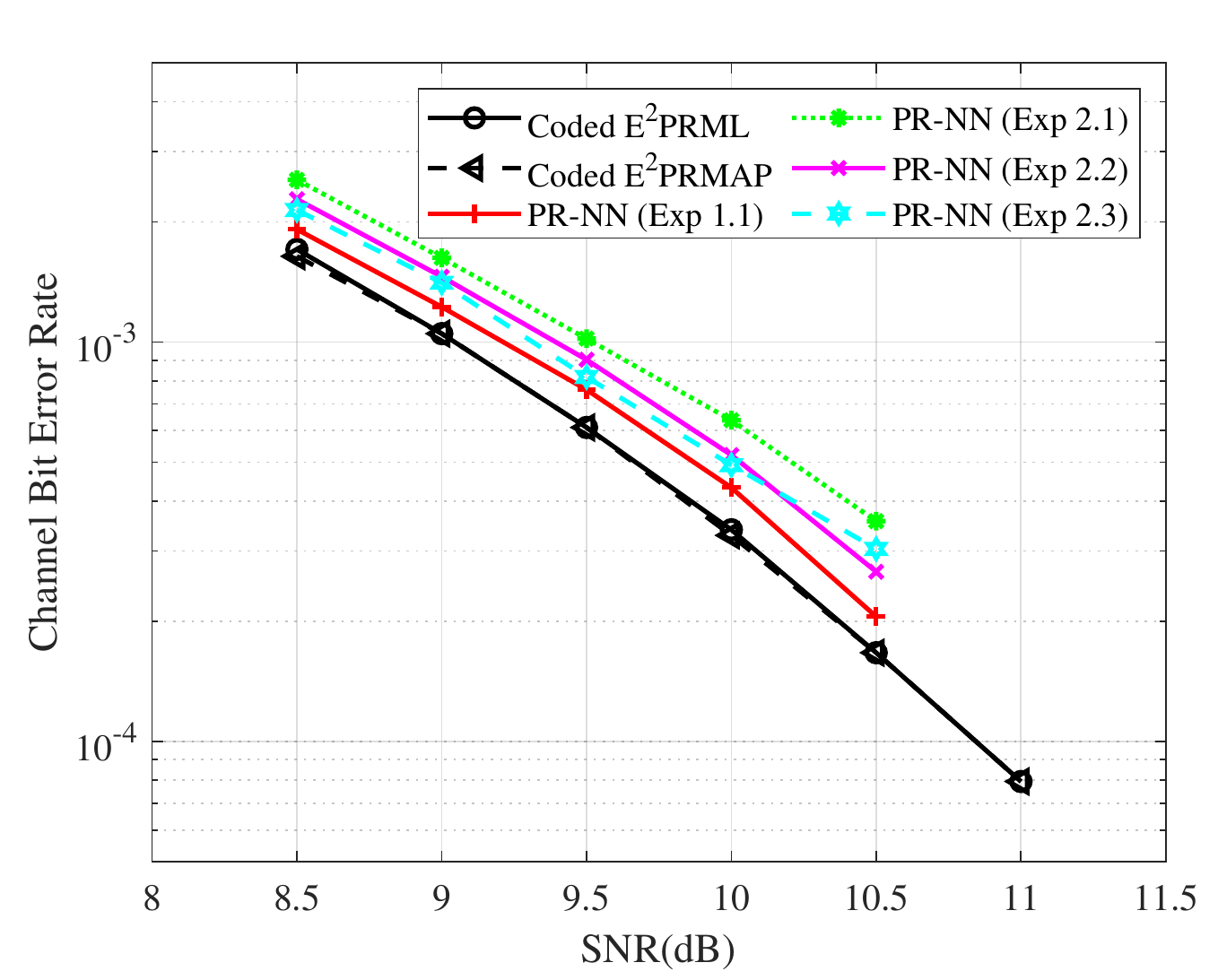}
		\caption{AWGN} 
		\label{fig::exp2allwhite}
	\end{subfigure}
	\begin{subfigure}[b]{0.45\columnwidth}
		\includegraphics[width=\linewidth]{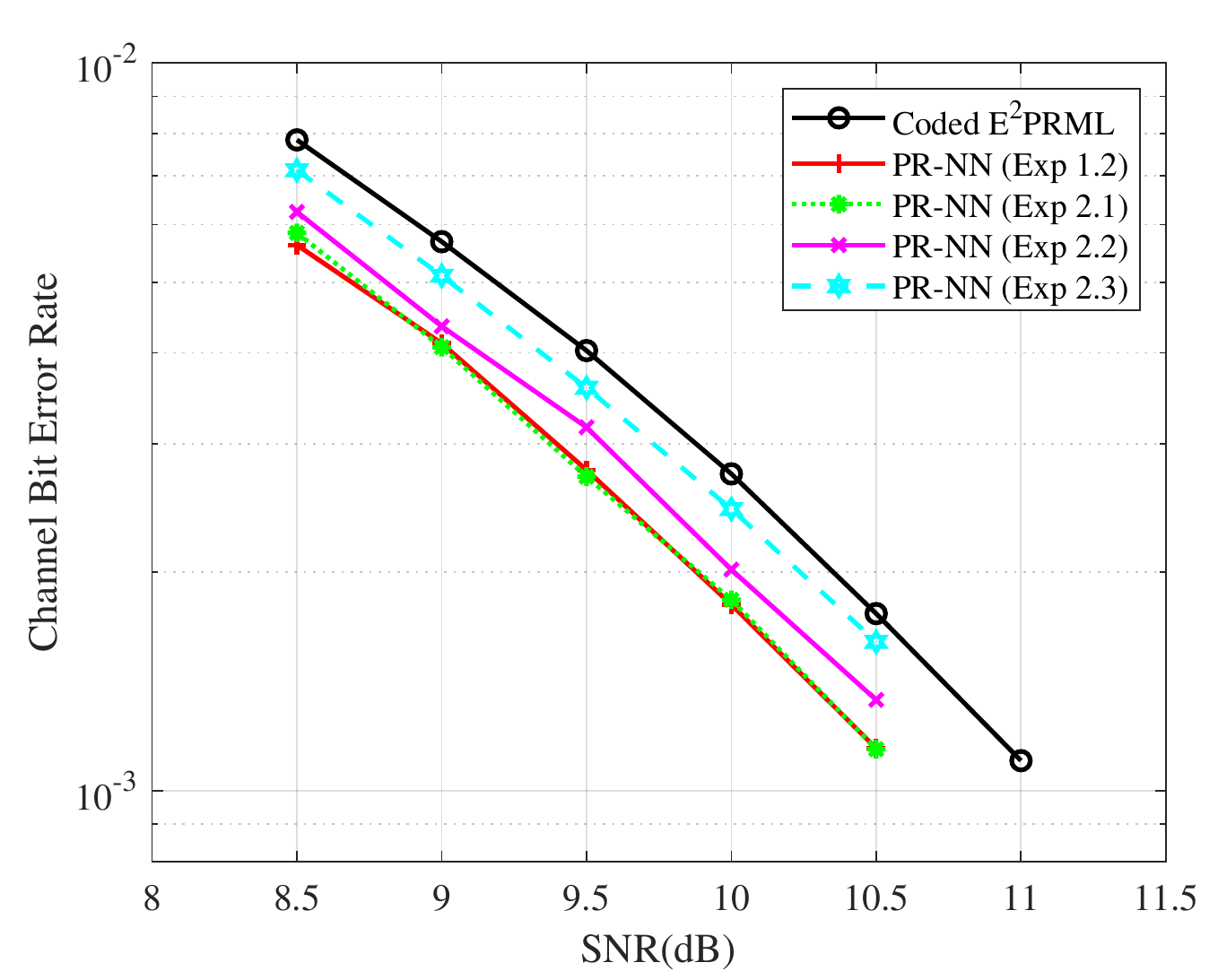}
		\caption{ACN ($PW_{50}/T_{c}=2.54$)} 
		\label{fig::exp2allcolor}
	\end{subfigure}
	\vspace{1ex}
	\caption{Scenario 2: Joint training with AWGN and ACN ($PW_{50}/T_{c}=2.54$).}
	\label{fig::exp2whitecolor}
	\vspace{-3ex}
\end{figure}

\textbf{Experiments 2.1, 2.2, 2.3}: First, the PR-NN detector is trained with both AWGN and ACN ($PW_{50}/T_{c}=2.54$). Experiments  2.1, 2.2, and 2.3 use different  relative batch sizes  for AWGN and ACN samples within the training set, as shown in Table~\ref{table::snrbtsize2}.  The simulation results are summarized in Fig.~\ref{fig::exp2whitecolor}. The solid red curve in Fig.~\ref{fig::exp2allwhite} represents the best performance in AWGN achieved by the network individually trained with AWGN (Experiment 1.1). When  trained with the combined datasets, the jointly trained PR-NNs suffer losses of 0.4dB, 0.2dB,  and 0.1dB, respectively, with respect to the network individually trained with ACN (Experiment 1.2), with the larger relative batch size  for AWGN giving the best performance.  On the other hand, as shown in Fig.~\ref{fig::exp2allcolor}, when we evaluate the jointly-trained PR-NN under ACN ($PW_{50}/T_{c}=2.54$), the performance losses are 0dB, 0.1dB, and 0.4dB, respectively, with increasing relative AWGN batch size.


These results suggest that  the best compromise in terms of robustness of performance is offered by the jointly trained PR-NN in Experiment 2.2, with losses of only 0.2dB in AWGN and 0.1dB in ACN.


\begin{figure}[t]
	\centering
	\begin{subfigure}[b]{0.45\columnwidth}
		\includegraphics[width=\linewidth]{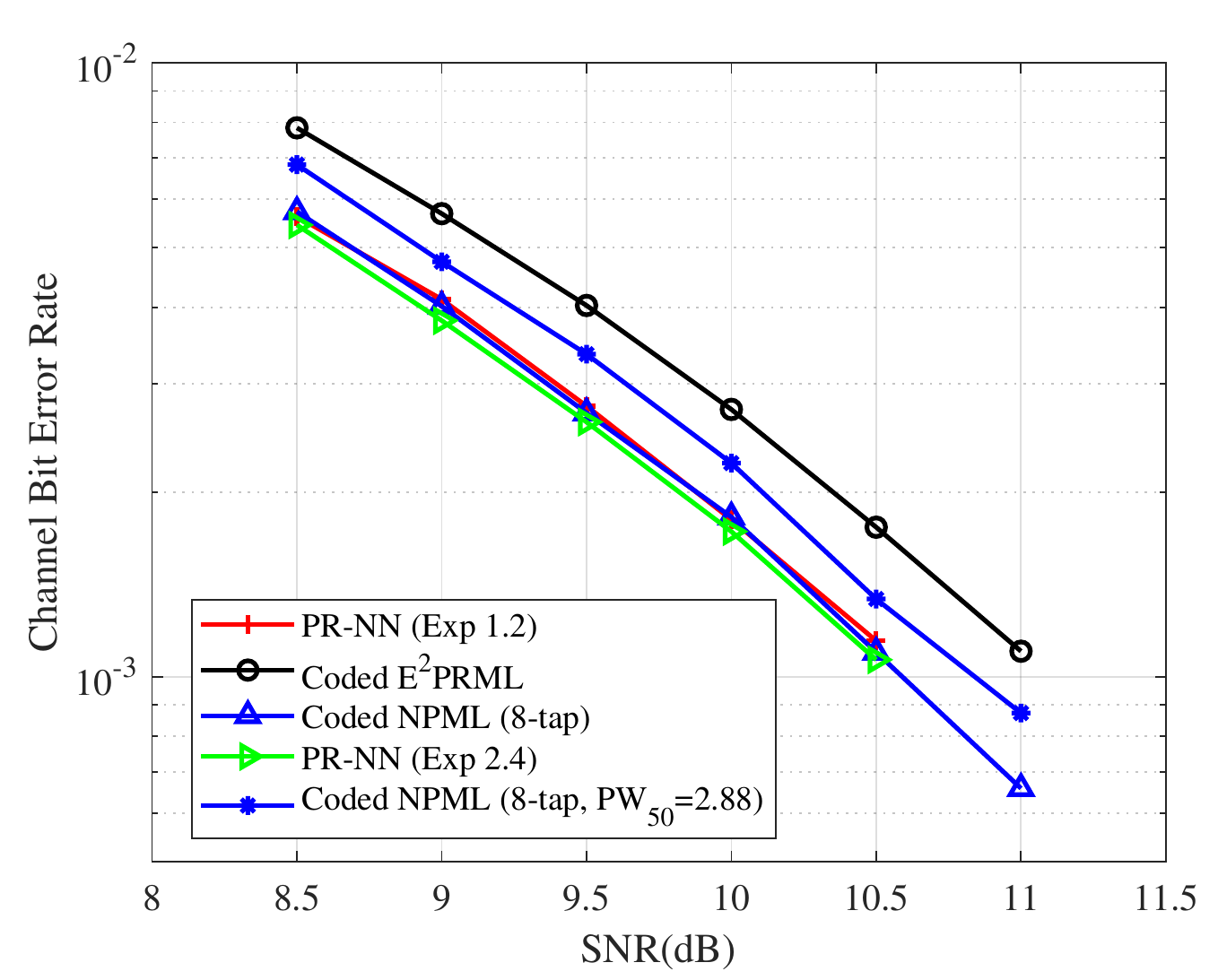}
		\caption{$PW_{50}/T_{c}=2.54$} 
		\label{fig::exp2allcolor254}
	\end{subfigure}
	\begin{subfigure}[b]{0.45\columnwidth}
		\includegraphics[width=\linewidth]{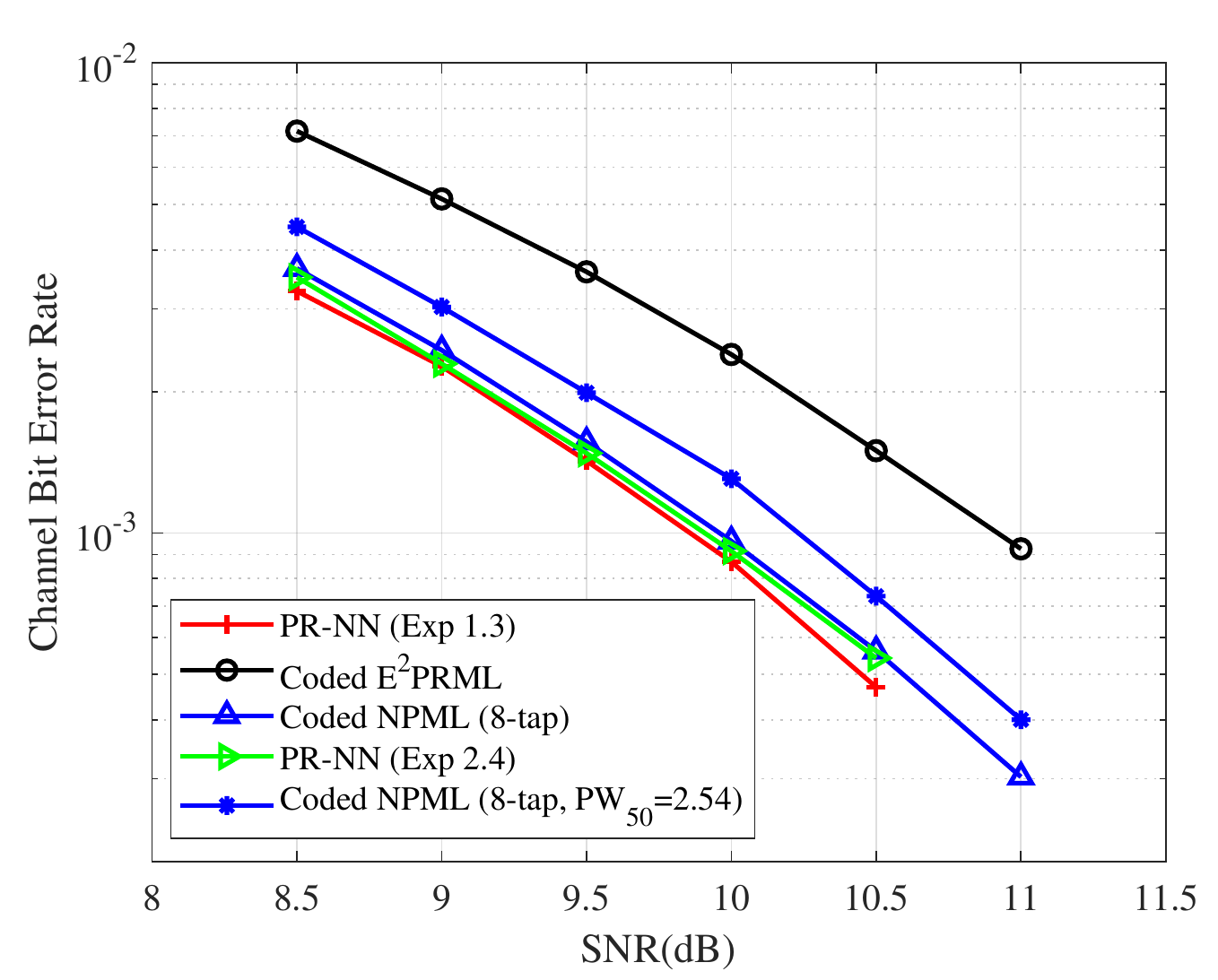}
		\caption{$PW_{50}/T_{c}=2.88$} 
		\label{fig::exp2allcolor288}
	\end{subfigure}
	\vspace{1ex}
	\caption{Scenario 2: Joint training with ACN ($PW_{50}/T_{c}=2.54$) and ACN ($PW_{50}/T_{c}=2.88$).}
	\label{fig::exp2color}
	\vspace{-3ex}
\end{figure}

\textbf{Experiment 2.4}: In this experiment, we explore the robustness of a jointly-trained PR-NN detector at different channel densities ($PW_{50}/T_{c}=2.54$ and $PW_{50}/T_{c}=2.88$). The training batch sizes are shown in Table~\ref{table::snrbtsize2}. 
The simulation results in Fig.~\ref{fig::exp2allcolor254} show that the resulting \mbox{PR-NN} detector matches the performance of the individually-trained network in Experiment 1.2. Moreover, as seen in Fig.~\ref{fig::exp2allcolor288}, the performance is only slightly worse than that of the network trained in Experiment 1.3. Thus, the jointly-trained network appears to offer adaptivity to different recording densities, which can arise from system variations in temperature and head flying height, or at different disk radii. 

Interestingly, the NPML detectors do not exhibit the same sort of robustness as the \mbox{PR-NN} detector. Fig.~\ref{fig::exp2allcolor254} shows that the NPML detector optimized for $PW_{50}/T_{c}=2.88$ experiences a performance loss of 0.2dB with respect to the  NPML detector properly optimized for $PW_{50}/T_{c}=2.54$. Similarly, we see in Fig.~\ref{fig::exp2allcolor288} that the NPML detector designed for $PW_{50}/T_{c}=2.54$ incurs a penalty of 0.3dB compared to the NPML detector designed for $PW_{50}/T_{c}=2.88$.


\subsection{Scenario 3: ``Realistic'' Equalized Lorentzian Channel}

In a ``realistic'' recording system modeled as an equalized  Lorentzian channel, the signal distortions include both colored noise and misequalization errors, as shown in~(\ref{eq::noise}). In this third set of experiments, we first compare the performance of  NPML detection  and PR-NN detection for such a system at two channel densities. We then assess the robustness of a PR-NN detector trained jointly for use at both densities. Note that in these experiments, the NPML detector designs take into account both colored noise and misequalization errors. 

\begin{figure}[t]
	\centering
	\begin{subfigure}[b]{0.45\columnwidth}
		\includegraphics[width=\linewidth]{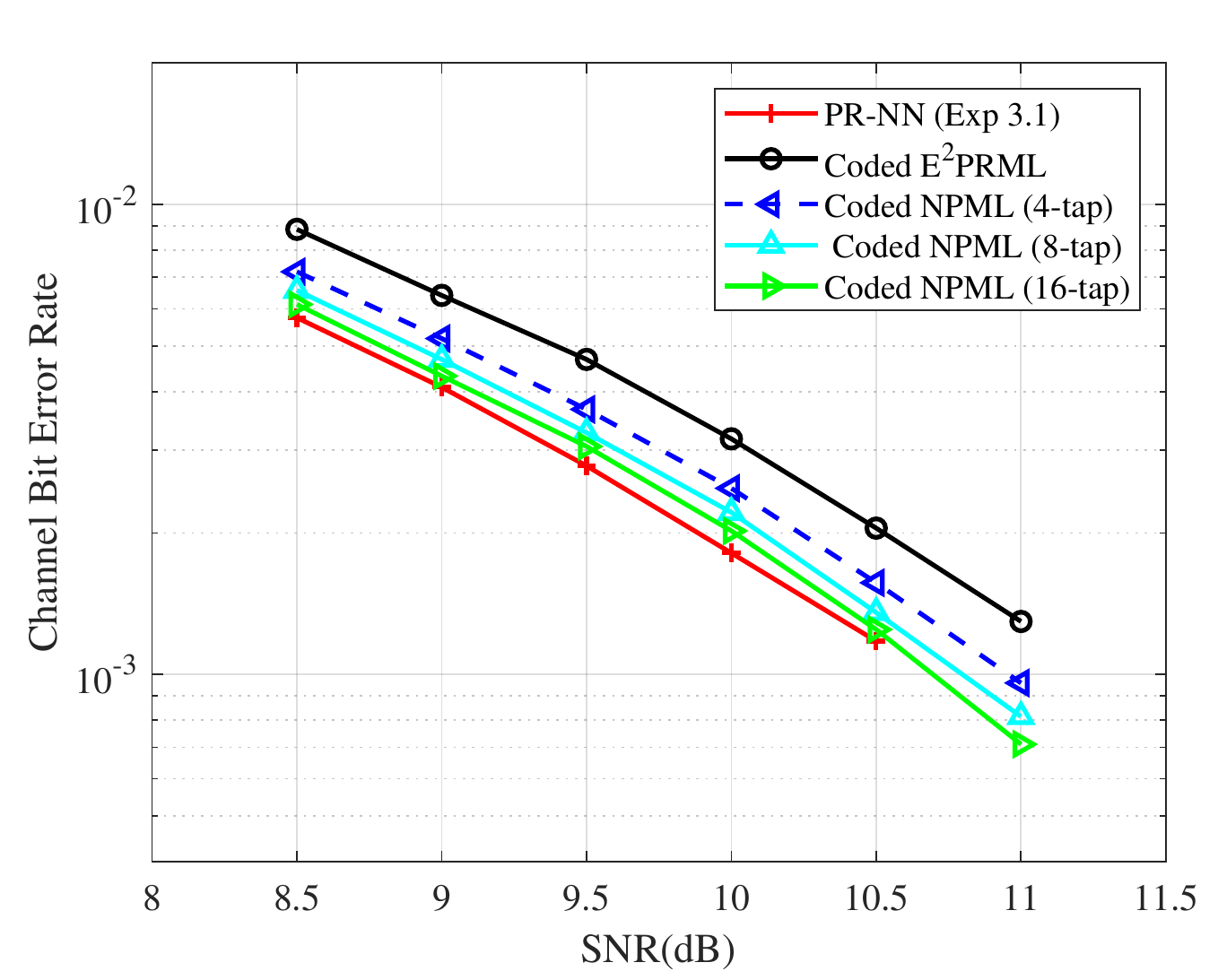}
		\caption{$PW_{50}/T_{c}=2.54$} 
		\label{fig::exp3254}
	\end{subfigure}
	\begin{subfigure}[b]{0.45\columnwidth}
		\includegraphics[width=\linewidth]{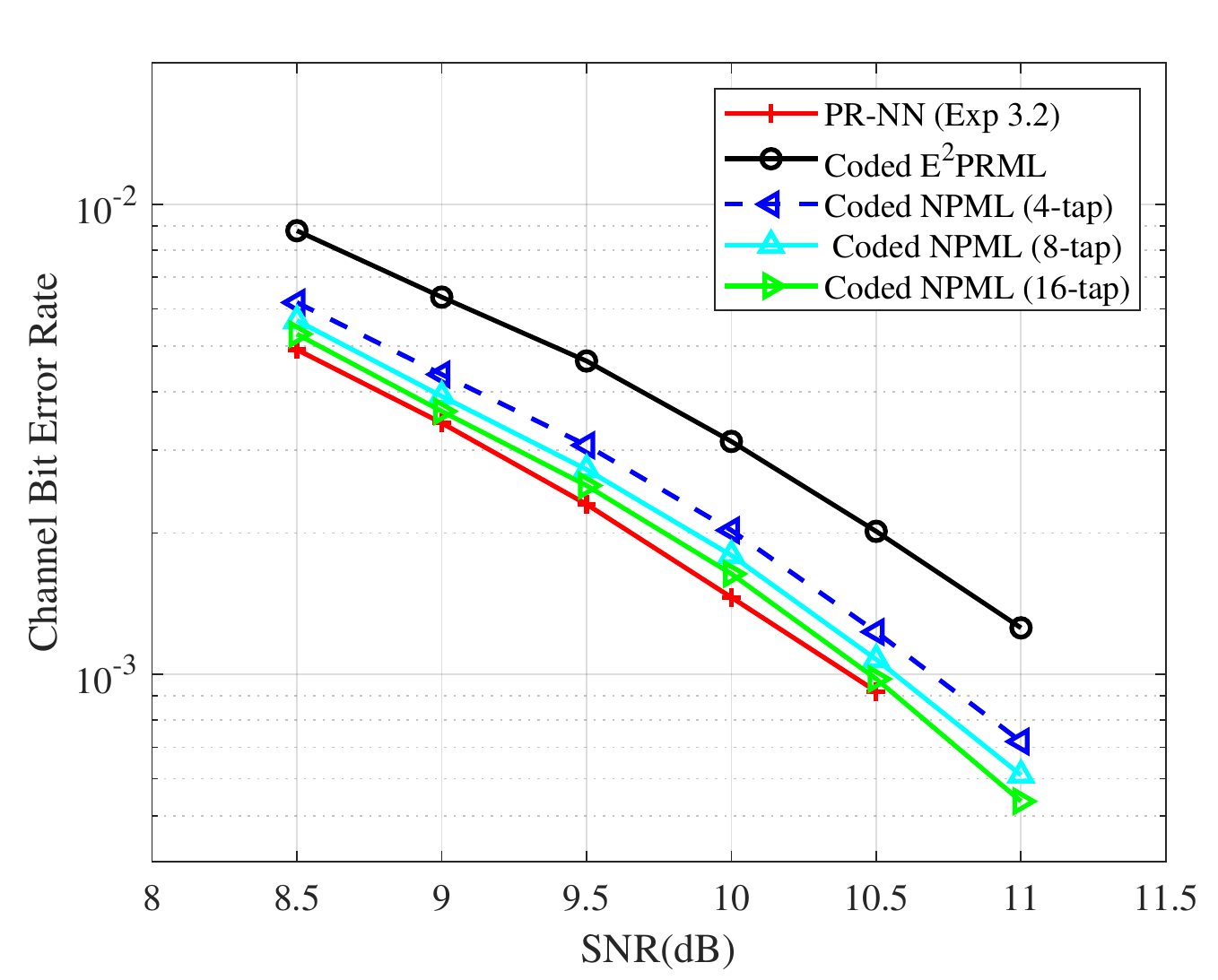}
		\caption{$PW_{50}/T_{c}=2.88$} 
		\label{fig::exp3288}
	\end{subfigure}
	\vspace{1ex}
	\caption{Scenario 3: Individual training with ``realistic'' datasets at  $PW_{50}/T_{c}=2.54$ and $PW_{50}/T_{c}=2.88$.}
	\label{fig::exp3}
	\vspace{-2ex}
\end{figure}

\textbf{Experiment 3.1}: The simulation results  for a PR-NN detectors trained individually at $PW_{50}/T_{c}=2.54$ are shown in Fig.~\ref{fig::exp3254}. The NPML detectors with $4$-tap, $8$-tap, and $16$-tap predictors have gains of 0.3dB, 0.4dB and 0.45dB, respectively, over coded E\textsuperscript{2}PRML. The \mbox{PR-NN} detector trained with the ``realistic'' channel  dataset achieves even slightly better  performance than the  $16$-tap NPML detector.


\textbf{Experiment 3.2}: In Fig~\ref{fig::exp3288}, we consider the channel with density $PW_{50}=2.88$. Here the gains of 
the NPML detectors with $4$-tap, $8$-tap, and $16$-tap predictors are 0.5dB, 0.6dB, and 0.7dB, respectively, over coded E\textsuperscript{2}PRML. The PR-NN detector trained with the corresponding dataset surpasses that of the $16$-tap NPML detector. 


\textbf{Experiment 3.3}: As in Experiment 2.4, we explore the adaptability of PR-NN detection to changes in recording density. 
The results obtained after training with a combined dataset of ``realistic'' equalized Lorentzian channel outputs for $PW_{50}=2.54$ and $PW_{50}=2.88$ are shown in Fig.~\ref{fig::exp3}. In Fig.~\ref{fig::exp3254}, corresponding to $PW_{50}=2.54$ , we see that the jointly-trained network essentially  matches the performance of the  individually trained network (Experiment 3.1), surpassing the $16$-tap NPML detector. Similarly,
Fig.~\ref{fig::exp3288} shows that the jointly-trained PR-NN detector preserves the performance of the network individually trained at $PW_{50}=2.88$ (Experiment 3.2). The robustness of PR-NN detection in this more realistic channel setting is thus confirmed. 


\begin{figure}[t]
	\centering
	\begin{subfigure}[b]{0.45\columnwidth}
		\includegraphics[width=\linewidth]{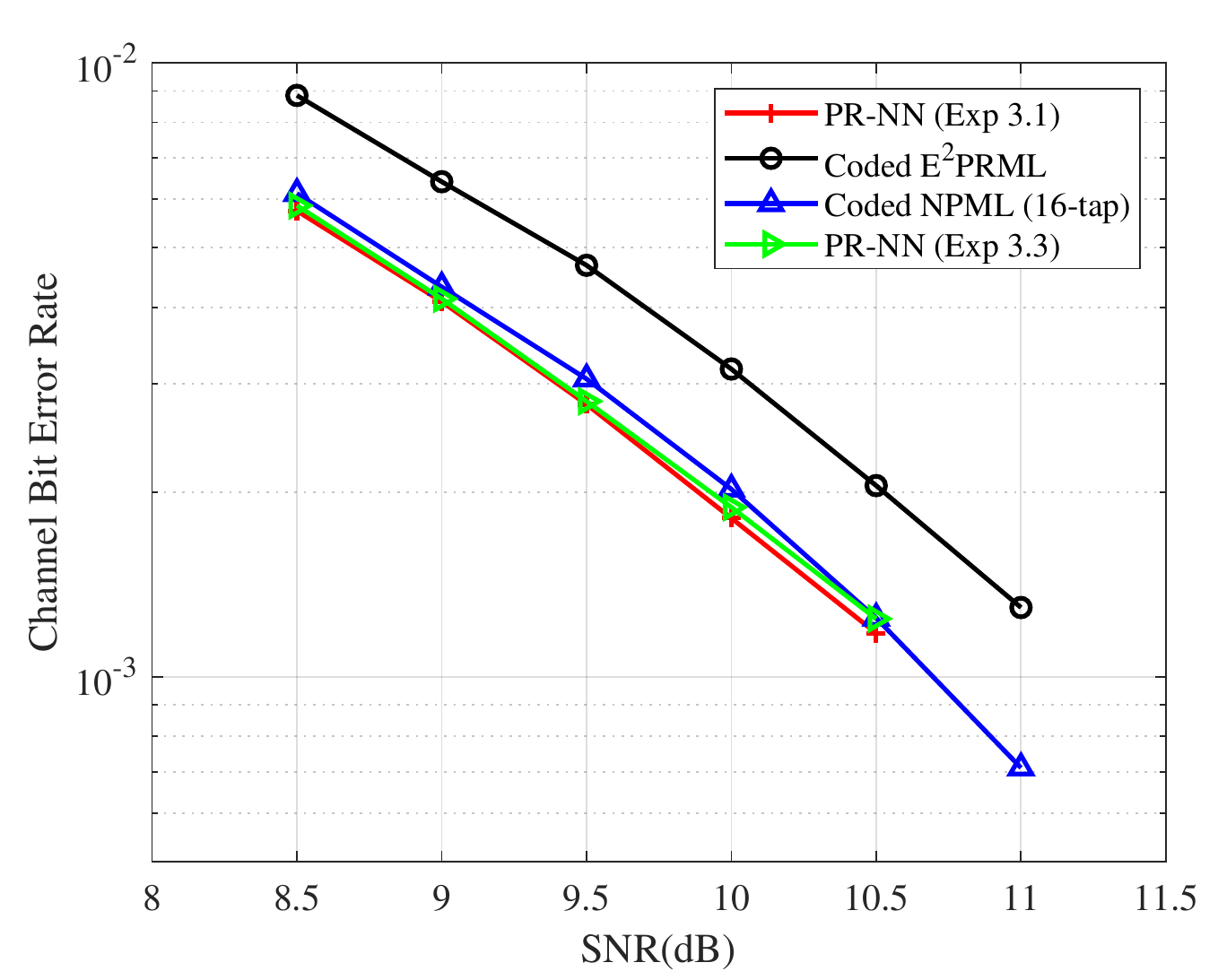}
		\caption{$PW_{50}/T_{c}=2.54$} 
		\label{fig::exp3all254}
	\end{subfigure}
	\begin{subfigure}[b]{0.45\columnwidth}
		\includegraphics[width=\linewidth]{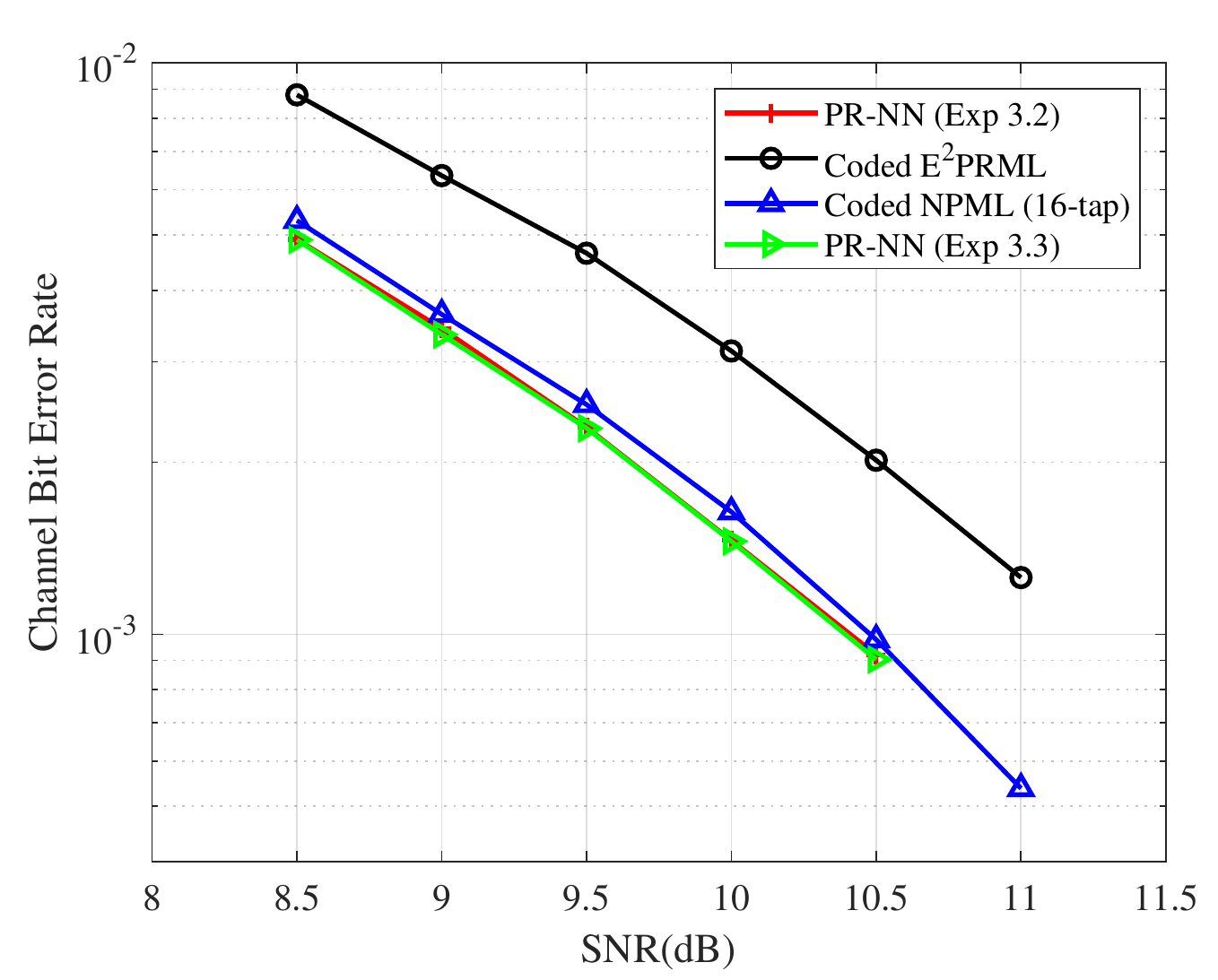}
		\caption{$PW_{50}/T_{c}=2.88$} 
		\label{fig::exp3all288}
	\end{subfigure}
	\vspace{1ex}
	\caption{Scenario 3: Joint training with ``realistic'' datasets for both $PW_{50}/T_{c}=2.54$ and $PW_{50}/T_{c}=2.88$.}
	\label{fig::exp3all}
	\vspace{-3ex}
\end{figure}

\subsection{Experimental Analysis}

Our results from Scenario 1 demonstrate that the PR-NN detection architecture can achieve performance close to Viterbi detection and NPML detection on   coded E\textsuperscript{2}PR4 channels in AWGN and ACN, respectively,  over a range of SNRs. 
In Scenario 2, we saw that a  PR-NN detector jointly trained for AWGN and ACN shows greater tolerance to ACN than the Viterbi detector, and retains comparable  performance in AWGN. When jointly trained in ACN corresponding to equalizers for two different channel densities, the PR-NN detector again exhibits robust performance over a range of SNRs. Finally, when evaluated on a more realistic equalized Lorentzian channel model with both ACN and misequalization errors, the PR-NN detectors designed individually for two channel densities surpass the performance of $16$-tap NPML detectors over a range of SNRs. The jointly-trained PR-NN detector maintains the performance of the individually-trained networks at both densities, displaying a robustness that the NPML detectors fail to offer.

The near-optimal performance and robustness of PR-NN can be explained as follows: 1) the RNN-based structure of PR-NN, which  exploits the time-sequential connections between cells, reflects the  nature of the signal generated by the ISI channel; 2) non-linear functions included in the RNN help PR-NN to model the effects of a variety of noise sources and distortions; and 3) the sliding-window evaluation process helps the block-wise PR-NN  architecture to detect the noisy outputs in streaming fashion, in analogy to the classical detection methods used in magnetic recording channels.

\section{Conclusion}

In this paper, we first formulated a magnetic recording channel model and reviewed three classical detectors. Then, we proposed PR-NN, an RNN-based detection approach for  coded partial-response channel models. The PR-NN detector processes the noisy outputs of the equalized recording channel in a block-streaming fashion, with computational complexity comparable to that of classical sequence detectors. Simulation results confirm the attractive performance of PR-NN when compared to classical detection algorithms and, moreover, demonstrate a robustness to different noise characteristics and channel densities that classical methods can not provide. 


\section*{Acknowledgment}
The authors would like to acknowledge helpful discussions with Joseph Soriaga.

\begingroup
\setstretch{1.41}

\endgroup
\end{document}